\documentclass[prd,preprintnumbers,nofootinbib]{revtex4} 
\usepackage{graphicx} 
\usepackage{amsmath}
\usepackage{amsfonts,amsbsy}
\usepackage{amssymb}
\usepackage{appendix}
\usepackage{slashed}

\def\be{\begin{equation}}
\def\ee{\end{equation}}
\def\bea{\begin{eqnarray}}
\def\eea{\end{eqnarray}}

\def\eq#1{{Eq.~(\ref{#1})}}
\def\fig#1{{Fig.~\ref{#1}}}
\def\fig#1{{Fig.~\ref{#1}}}

\begin{document}

\title{\bf Di-photon "Ridge"  in p+p and p+A collisions at RHIC and the LHC}


\author{Alex Kovner$^{1}$ and Amir H. Rezaeian$^{2,3}$}
\affiliation{
$^1$Dept. of Physics, University of Connecticut, High, Storrs, CT 06269, USA\\
$^2$Departamento de F\'\i sica, Universidad T\'ecnica
Federico Santa Mar\'\i a, Avda. Espa\~na 1680,
Casilla 110-V, Valparaiso, Chile\\
$^3$  Centro Cient\'\i fico Tecnol\'ogico de Valpara\'\i so (CCTVal), Universidad T\'ecnica
Federico Santa Mar\'\i a, Casilla 110-V, Valpara\'\i so, Chile
}

\begin{abstract}
We obtain prompt di-photon cross-section in proton-nucleus collisions in Hamiltonian light-cone approach within a hybrid approximation, treating the projectile proton in the parton model and  the target nucleus in the Color-Glass-Condensate approach. We study in details the di-photon correlations in quark-nucleus and proton-nucleus collisions at RHIC and the LHC. We show that the single fragmentation di-photon produces the away side correlations peak, and the double fragmentation component of prompt di-photon is responsible for the near-side peak, and the long-range in rapidity near-side azimuthal collimation, the so-called "ridge" structure. We study the transverse momentum, density and energy dependence of the di-photon ridge  and show that it strongly depends on the kinematics and saturation dynamics. We show that while di-photon ridge exists at the LHC in quark-nucleus collisions, the effect disappears in proton-nucleus collisions at the LHC. At RHIC the ridge-type structure persists at low transverse momenta of di-photon even in proton-nucleus collisions. We argue that di-photon correlation measurments in p+A collisions can help to discriminate among models and understand the true origin of the observed di-hadron ridge in p+A collisions.  We also show that in addition to the ridge structure, prompt di-photon correlation also exhibits some distinct novel features, including the emergence of away side double-peak structure at intermediate transverse momenta. 

\end{abstract}

\maketitle

\section{Introduction}
Recent experimental observations of two hadrons correlations in the relative azimuthal  angle $\Delta \phi=\phi_1-\phi_2$ and in the pseudorapidity separation $\Delta\eta=\eta_1-\eta_2$ in high-multiplicity proton-proton (p+p) \cite{exp-pp} and proton(deuteron)-nucleus (p+A) \cite{exp-pa1,exp-pa2,exp-pa3,exp-pa4,exp-pa5,exp-pa6} collisions, show a great deal of similarity to that measured in semi peripheral nucleus-nucleus collisions \cite{exp-hic}. In particular, the discovery of the so-called ridge phenomenon, namely the near-side ($\Delta \phi\approx 0$)  di-hadron correlations which extend to a large pseudorapidity separation $\Delta\eta$, in high-multiplicity events selection in both p+p and p(d)+A collisions at the LHC and RHIC \cite{exp-pp,exp-pa1,exp-pa2,exp-pa3,exp-pa4,exp-pa5,exp-pa6}, triggered an on-going debate about the underlying dynamics of high-multiplicity events. A similar ridge-type structure has also been observed in heavy-ion collisions at RHIC and the LHC and is understood as a phenomenon related to hydrodynamical behavior of the quark-gluon-plasma, see e.g Refs.\,\cite{exp-geo1,exp-geo2}. The hydrodynamic approach, based on the final-state interaction provides good description \cite{hydro-1,hydro-2} of the high multiplicity p+p data as well. Nevertheless, no compelling argument has yet been given that a small system like the one produced in p+p collisions, an order of magnitude  smaller than heavy-ion collisions, should exhibit a hydro type behavior.   On the other hand, the initial-state Color-Glass-Condensate (CGC) approach, the effective field theory of low-x partons in the hadronic or nuclear wavefunctions, also provides a qualitatively good description of the same data \cite{ridge0,ridge1,ridge2,ridge3,ridge4,ridge5,ridge6,ridge7,ridge8,ridge9,ridge10}. It is thus fair to say that the true physical origin of the ridge in p+p and p+A collisions is still unknown, see also Refs. \cite{ridge-o1,ridge-o2,ridge-o3,ridge-o4,ridge-o5,ridge-o6,ridge-o7,ridge-o8,v2-amir}.

The mechanism of prompt di-photon (photons produced in hard scatterings, not from hadron decay) production \cite{di-photon} in the CGC/saturation framework is quite different than that of di-hadron production \cite{di-hadron}.  In the CGC picture, hadrons are produced in two stages: first soft gluons (or gluon mini-jets) are put on-shell from the projectile wave function by directly scattering on a saturated target, and subsequently gluon mini-jets decay to hadrons via a final-state hadronization process. The first stage is theoretically under control  while the last stage of hadronization can be only treated phenomenologically in the small-x  framework using the gluon fragmentation functions \cite{hybrid}.  In contrast to gluons, prompt photons do not scatter on the target gluon field, but rather decohere from the projectile wavefunction due to scattering of the quarks. Moreover, prompt photons are free from the final-state hadronization processes, and possible initial state-final state interference effects.  Therefore, prompt photons can be a powerful probe of initial-state effects, and in particular the prompt di-photon correlations can provide vital information about the intrinsic correlations of partons in the hadronic and nuclear wavefunctions. Such measurments can also shed light on the origin of the observed ridge phenomenon in di-hadron production and help us understand  whether it is the initial or final state effects that play dominant role in formation of the ridge collimation in p+p and p+A collisions.

Note also that unlike description of di-hadron correlations \cite{hybrid,di-hadron} which requires the knowledge of correlators of a higher number of Wilson lines, the di-photon production cross section (at least in leading order in $\alpha_s$) depends only on the dipole amplitude \cite{di-photon}, which is the best understood observable in terms of the high energy evolution.  Therefore, within the CGC framework, the theoretical understanding of observables necessary to describe di-photon production is more robust compared to di-hadron production.

The main aim of this paper is to investigate prompt di-photon  azimuthal angular correlations  in p+p and p+A collisions at RHIC and the LHC. We concentrate on relatively forward rapidities and employ the hybrid approximation \cite{hybrid} within the CGC approach.  In Ref.\,\cite{di-photon}, we calculated di-photon cross-section in the hybrid approximation  at the leading-order in the CGC framework employing the soft approximation. Since in the CGC approach the scattered quark does not change its longitudinal momentum while propagating through the target,  the soft approximation amounts to the assumption that the transverse momenta of the produced photons are smaller that the typical transverse momentum of exchanges with the target. Although the soft approximation greatly simplifies calculations, it is not applicable in a significant range of interesting kinematics in high-energy collisions. Here, for the first time, we provide the full calculation of prompt di-photon production in p+A collisions  at leading-order in the CGC framework without resorting to any further approximation.  

We show that prompt di-photon in high-energy collisions exhibit  ridge-type structure in quark-nucleus collisions albeit  the physics of these correlations is significantly different than that of hadronic correlations discussed previously in the CGC framework \cite{ridge0,ridge1,ridge2,ridge3,ridge4,ridge5,ridge6,ridge7,ridge10}. In particular the correlations decrease for moderate rapidity separation between the two photons ($\sim 3$ units).   We investigate the energy, rapidity, transverse momentum, and density dependence of prompt di-photon correlations and show that the di-photon correlations have many novel features. We attribute the intrinsic  azimuthal collimation (the ridge) in prompt di-photon to the double-fragmentation contribution at relatively high transverse momentum while the away-side peak to single fragmentation contribution coming mainly from relatively low transverse momentum. We also show that the degree of manifestation of these correlations in the final state is strongly influenced by the gluon saturation dynamics.

This paper is organized as follows:  In Sec. II we introduce the main formalism for calculating the cross-section of inclusive prompt di-photon production in high-energy proton-proton and proton-nucleus collisions in the CGC framework and obtain the cross-section within the Hamiltonian light-cone approch at leading-order.  In Sec. III we present the results of numerical calculations for correlations of prompt di-photon production in p+p and p+A collisions at RHIC and the LHC. We summarize our main results in Sec. IV. The details of the calculation are given in the Appendix.


\section{Inclusive prompt di-photon production in the color-glass-condensate approach }
\begin{figure}[t]                                       
                                \includegraphics[width=14 cm] {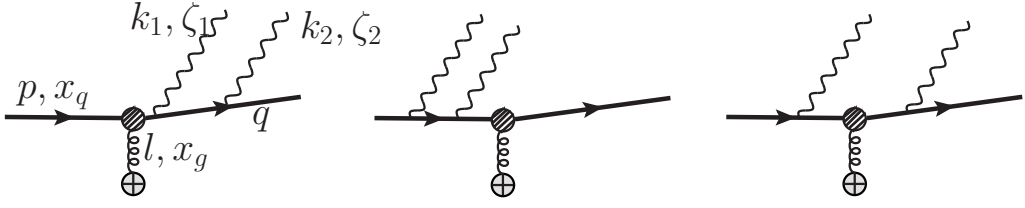}                                  
\caption{Typical CGC diagrams contributing to the semi-inclusive prompt di-photon-quark production at leading-order in quark-nucleus collisions. The crossed blob denotes the interaction of the projectile quark to all orders with the strong background field of the target nucleus. }
\label{f-di-cgc}
\end{figure}

  In the CGC approach the quantum corrections enhanced by large logarithms of $1/x$ are systematically re-summed incorporating high gluon density effects at low x and for large nuclei \cite{sg,mv,jimwlk,bk}. The inclusive prompt di-photon production $h+A\to \gamma_1\,+\,\gamma_2\,+\, X$ in high-energy dilute-dense scatterings  was recently calculated in the CGC approach \cite{di-photon} in the soft approximation where  a dilute projectile hadron (h) interacts coherently with a dense target $A$ and produces two prompt photons $\gamma_1$ and $\gamma_2$. In the leading order approximation, at forward rapidity,  a valence quark of the projectile hadron emits two photons via Bremsstrahlung and the produced di-photon+jet is then put on shell by interacting coherently over the whole longitudinal extent of the target, see \fig{f-di-cgc}.

Here we  derive the expression for the prompt di-photon production cross section off a valence quark in the wave function formalism without resorting to the soft approximation.  
The relevant part of the light front Hamiltonian  which contains the photon-quark interaction is

\begin{eqnarray}
H_I&=&\frac{1}{2}g\int_{p^+,k^+>0}\frac{1}{\sqrt{2q^+}}\bigg\{\frac{2p^++q^+}{p^++q^+}\bigg[\frac{p_i}{p^+}-\frac{q_i}{q^+}\bigg]\delta_{ij}\delta_{s',s}-i\epsilon^{ij}\sigma^3_{s',s}\frac{q^+}{p^++q^+}\bigg[\frac{p_i}{p^+}-\frac{q_i}{q^+}\bigg]\bigg\} \nonumber \\ 
&\times&\bigg[b_{s'}^{\alpha}(p^++q^+,p+q)b_s^{\dagger\alpha}(p^+,p)a^\dagger_j(q^+,q)+d_{s'}^{\dagger\alpha}(p^+,p)d_s^{\alpha}(p^++q^+,p+q)a^\dagger_j(q^+,q)\bigg] +h.c.
\end{eqnarray}
Here  $\{\alpha,s,i\}$ are color, spin  and transverse Lorentz indices respectively, $(p,q)$ are transverse momenta and $b^\alpha_s$, $d^\alpha_s$ and $a_i$ are the quark, antiquark and photon annihilation operators respectively.
The terms we omitted here  only contribute to virtual corrections for the one quark state and are unimportant in the leading order approximation.
The one quark state dressed by the photons to order $g^2$ has a general form
\begin{eqnarray}\label{D}
|p,s\rangle_D&=&|p,s\rangle +\int_{q}\sum_{i,s'}B(p-q,s,s';q,i)|p-q,s'; q,i\rangle\nonumber \\ 
&+&\int_{q_1,q_2}\sum_{i,j,s"}C(p-q_1-q_2,s,s";q_1,i,q_2,j)|p-q_1-q_2,s";q_1,i,q_2,j\rangle.\ 
\end{eqnarray}
Here the momenta have both transverse and longitudinal component, and momenta $q_i$ refer to the momenta of photons in the wave function. The coefficient functions $B$ and $C$ will be determined from perturbation theory. The function $B$ is of order $g$, while $C$ is of order $g^2$.
The outgoing state after scattering on the target is
\begin{eqnarray}
|p,s\rangle_{out}&=&\int_lM(l)\Big[|p+l,s\rangle+\int_{q}\sum_{i,s'}B(p-q,s,s';q,i)|p+l-q,s'; q,i\rangle\nonumber\\
&+&\int_{q_1,q_2}\sum_{i,j,s"}C(p-q_1-q_2,s,s";q_1,i,q_2,j)|p+l-q_1-q_2,s";q_1,i,q_2,j\rangle\Big],\
\end{eqnarray}
where $M(l)$ is the probability amplitude to transfer momentum $l$ from the target to the quark. The momentum transfer from the target $l$ is assumed to be purely transverse.
As usual in this formalism we rewrite the outgoing state in the dressed quark basis
\begin{eqnarray}
&&|p,s\rangle_{out}=\int_lM(l)\Big[|p+l,s\rangle_D +\int_{q}\sum_{i,s'}[B(p-q,s,s';q,i)-B(p-q+l,s,s';q,i)]|p+l-q,s'; q,i\rangle\nonumber\\
&&+\int_{q_1,q_2}\sum_{i,j,s"}[C(p-q_1-q_2,s,s";q_1,i,q_2,j)-C(p+l-q_1-q_2,s,s";q_1,i,q_2,j)]|p+l-q_1-q_2,s";q_1,i,q_2,j\rangle\Big]. \nonumber\\
\end{eqnarray}
In this expression the one quark, one gluon state is still bare. To express it in terms of the dressed quark we need to use again \eq{D}, but only to order $g$ this time,
\begin{equation}
|p+l-q_1,s'; q_1,i\rangle_D=|p+l-q_1,s'; q_1,i\rangle+\int_{q_2}\sum_{j,s"}B(p+l-q_1-q_2,s',s";q_2,j)|p+l-q_1-q_2,s"; q_1,i,q_2,j\rangle. 
\end{equation}
The dressing corrections to the two photon state do not have to be taken into account, since they lead to corrections of at least a power  $g^3$ in the outgoing quark wave function. Thus the final expression for the outgoing state in terms of the "dressed" eigenstate of the Hamiltonian to order $g^2$ is
\begin{eqnarray}
&&|p,s\rangle_{out}=\int_lM(l)\Big[|p+l,s\rangle_D
+\int_{q}\sum_{i,s'}[B(p-q,s,s';q,i)-B(p-q+l,s,s';q,i)]|p+l-q,s'; q,i\rangle_D\nonumber\\
&&+\int_{q_1,q_2}\sum_{i,j,s"}\Big\{[B(p-q_1,s,s';q_1,i)-B(p+l-q_1,s,s';q_1,i)]B(p+l-q_1-q_2,s',s";q_2,j)\nonumber\\
&&+[C(p-q_1-q_2,s,s";q_1,i,q_2,j)-C(p+l-q_1-q_2,s,s";q_1,i,q_2,j)]\Big\}|p+l-q_1-q_2,s";q_1,i,q_2,j\rangle_D\Big].\ 
\end{eqnarray}
Let us define for convenience the amplitude
\begin{eqnarray}\label{F}
F(p,l,q_1,q_2,ijss")&\equiv& C(p-q_1-q_2,s,s";q_1,i,q_2,j)-C(p+l-q_1-q_2,s,s";q_1,i,q_2,j)\nonumber\\
&+&[B(p-q_1,s,s';q_1,i)-B(p+l-q_1,s,s';q_1,i)]B(p+l-q_1-q_2,s',s";q_2,j).\
\end{eqnarray}
The inclusive two photon cross section is (up to normalization) given by the following expectation value in the outgoing state
\begin{equation}
\frac{d\sigma^{qA\to \gamma(k_1)\gamma(k_2) X }}{d ^2k_1dk_1^+d^2k_2dk_2^+}\propto\langle out| a^\dagger_i(k_1)a^\dagger_j(k_2)a_i(k_1)a_j(k_2)|out\rangle.
\end{equation}
Additionally we average over the spin and the color index of the incoming quark.
The result is
\begin{eqnarray}\label{sigma}
\frac{d\sigma^{qA\to \gamma(k_1)\gamma(k_2) X }}{d ^2k_1dk_1^+d^2k_2dk_2^+}&=&\frac{1}{2} \sum_{s,s',i,j}\int_lN(l)\Big[F^*(0,l,k_1,k_2,i,j,ss')+F^*(0,l,k_2,k_1,j,i,ss')\Big] \nonumber\\
&\times&\Big[F(0,l,k_1,k_2,i,j,ss')+F(0,l,k_2,k_1,j,i,ss')\Big],\
\end{eqnarray}
where $N(l)=M^*(l)M(l)$ is the dipole scattering amplitude (see Sec. III).
Our next step is to calculate the  functions $B$ and $C$.
We use the general second order perturbative expressions for perturbation $V$.
\begin{equation}
|n\rangle_D=|n\rangle+\sum_{k\ne n}|k\rangle \frac{V_{kn}}{E_n-E_k}+\sum_{k\ne n, l\ne n}|k\rangle\frac{V_{kl}V_{ln}}{(E_n-E_l)(E_n-E_l)}+\dots,
\end{equation}
where we have omitted the terms involving diagonal matrix elements of $V$, which in our case vanish.
This gives the following explicit expressions
\begin{eqnarray}
&&B(p-q,s,s',q,i)=\frac{g}{\sqrt{2q^+}}\frac{1}{\frac{p^2}{2p^+}-\frac{(p-q)^2}{2(p^+-q^+)}-\frac{q^2}{2q^+}}\Big[\frac{(p-q)_j}{p^+-q^+}-\frac{q_j}{q^+}\Big]\Big\{\frac{2p^+-q^+}{p^+}\delta_{ji}\delta_{ss'}+i\frac{q^+}{p^+}\epsilon_{ji}\sigma^3_{ss'}\Big\}\nonumber\\
&&=\frac{g}{\sqrt{2q^+}}\frac{1}{\frac{p^2}{2p^+}-\frac{(p-q)^2}{2(p^+-q^+)}-\frac{q^2}{2q^+}}[q^+p_j-p^+q_j]\Big\{\left[\frac{1}{p^+q^+}+\frac{1}{(p^+-q^+)q^+}\right]\delta_{ji}\delta_{ss'}+i\frac{1}{p^+(p^+-q^+)}\epsilon_{ji}\sigma^3_{ss'}\Big\}, 
\end{eqnarray}
\begin{eqnarray}
&&C(p-q_1-q_2, s,s',q_1,i,q_2,j)=\frac{g^2}{\sqrt{4q_1^+q_2^+}}\left[\frac{1}{\frac{p^2}{2p^+}-\frac{(p-q_1)^2}{2(p^+-q_1^+)}-\frac{q_1^2}{2q_1^+}}\right]\left[\frac{1}{\frac{p^2}{2p^+}-\frac{(p-q_1-q_2)^2}{2(p^+-q_1^+-q_2^+)}-\frac{q_1^2}{2q_1^+}-\frac{q_2^2}{q_2^+}}\right]\nonumber\\
&&\times(q^+_1p_l-p^+q_{1l})\left[\left(\frac{1}{p^+q^+_1}+\frac{1}{q_1^+(p^+-q_1^+)}\right)\delta_{li}\delta_{ss''}
+\frac{i}{p^+(p^+-q_1^+)}\epsilon_{li}\sigma^3_{ss''}\right]\nonumber\\
&&\times(q_2^+(p_k-q_{1k})-(p^+-q^+_1)q_{2k})\nonumber\\
&&\times\left[\left(\frac{1}{q_2^+(p^+-q^+_1)}+\frac{1}{q_2^+(p^+-q_1^+-q_2^+)}\right)\delta_{kj}\delta_{s''s'}+\frac{i}{(p^+-q_1^+)(p^+-q_1^+-q_2^+)}\epsilon_{kj}\sigma^3_{s''s'}\right].\nonumber\\
\end{eqnarray}
In order to simplify the equations we rewrite these equation in terms of the energy ratios $\zeta_i$ defined as
\begin{equation}
q^+=\zeta p^+;\ q_1^+=\zeta_1p^+; \ q_2^+=\zeta_2 p^+, 
\end{equation}
The result is
\begin{equation}
B(p-q,s,s',q,i)=g\sqrt{\frac{2\zeta}{p^+}}\frac{1}{p^2\zeta(1-\zeta)-(p-q)^2\zeta-q^2(1-\zeta)}(\zeta p_j-q_j)
\left[\frac{2-\zeta}{\zeta}\delta_{ji}\delta_{ss'}-i\epsilon_{ji}\sigma^3_{ss'}\right].
\end{equation}
\begin{eqnarray}
&&C(p-q_1-q_2, s,s',q_1,i,q_2,j)=g^2\sqrt{\frac{4\zeta_1\zeta_2}{(p^+)^2}}\frac{\zeta_1}{1-\zeta_1}\frac{1}{p^2\zeta_1(1-\zeta_1)-(p-q_1)^2\zeta_1-q_1^2(1-\zeta_1)}
\big(\zeta_1 p_l-q_{1l}\big)\nonumber\\
&&\times \left[\frac{2-\zeta_1}{\zeta_1}\delta_{li}\delta_{ss'}-i\epsilon_{li}\sigma^3_{ss'}\right]\frac{1}{p^2\zeta_1\zeta_2(1-\zeta_1-\zeta_2)-(p-q_1-q_2)^2\zeta_1\zeta_2-q_1^2(1-\zeta_1-\zeta_2)\zeta_2-q_2^2(1-\zeta_1-\zeta_2)\zeta_1}\nonumber\\
&&\times\big(\zeta_2(p-q_1)_k-(1-\zeta_1)q_{2k}\big)\left[\frac{2-2\zeta_1-\zeta_2}{\zeta_2}\delta_{kj}\delta_{s''s'}-i\epsilon_{kj}\sigma^3_{s''s'}\right]. 
\end{eqnarray}
After a simple algebra in the denominators we find
\begin{equation}
B(p-q,s,s',q,i)=-g\sqrt{\frac{2\zeta}{p^+}}\frac{1}{(\zeta p-q)^2}(\zeta p_j-q_j)
\left[\frac{2-\zeta}{\zeta}\delta_{ji}\delta_{ss'}-i\epsilon_{ji}\sigma^3_{ss'}\right].
\end{equation}
\begin{eqnarray}
&&C(p-q_1-q_2, s,s',q_1,i,q_2,j)=g^2\sqrt{\frac{4\zeta_1\zeta_2}{(p^+)^2}}\frac{\zeta_1}{1-\zeta_1}\frac{\big(\zeta_1 p_l-q_{1l}\big)}{(\zeta_1p-q_1)^2}
\left[\frac{2-\zeta_1}{\zeta_1}\delta_{li}\delta_{ss'}-i\epsilon_{li}\sigma^3_{ss'}\right]\\
&&\times\frac{(\zeta_2p-q_2)_k+(\zeta_1q_2-\zeta_2q_1)_k}{\zeta_2(\zeta_1p-q_1)^2  +\zeta_1(\zeta_2p-q_2)^2 -(\zeta_2q_1-\zeta_1q_2)^2}
\left[\frac{2-2\zeta_1-\zeta_2}{\zeta_2}\delta_{kj}\delta_{s''s'}-i\epsilon_{kj}\sigma^3_{s''s'}\right].\nonumber
\end{eqnarray}
We can also obtain the following useful identities, 
\begin{equation}
B(-q,s,s',q,i)-B(l-q,s,s',q,i)=-g\sqrt{\frac{2\zeta}{p^+}}\left[-\frac{q_j}{q^2}-\frac{\zeta l_j-q_j}{(\zeta l-q)^2}\right]\left[\frac{2-\zeta}{\zeta}\delta_{ji}\delta_{ss'}-i\epsilon_{ji}\sigma^3_{ss'}\right].
\end{equation}
\begin{eqnarray}
&&\left[B(-q_1,s,s',q_1,i)-B(l-q_1,s,s',q_1,i)\right]B(l-q_1-q_2,s',s";q_2,j)=\nonumber\\
&&g^2\frac{\sqrt{4\zeta_1\zeta_2}}{p^+}\left[-\frac{q_{1l}}{q_1^2}-\frac{\zeta_1 l_l-q_{1l}}{(\zeta_1 l-q_1)^2}\right]\frac{(\zeta_2l-q_2)_k+(\zeta_1q_{2}-\zeta_2q_1)_k}{\Big((\zeta_2l-q_2)+(\zeta_1q_2-\zeta_2q_1)\Big)^2}\nonumber\\
&&\times
\left[\frac{2-\zeta_1}{\zeta_1}\delta_{li}\delta_{ss'}-i\epsilon_{li}\sigma^3_{ss'}\right]\left[\frac{2(1-\zeta_1)-\zeta_2}{\zeta_2}\delta_{kj}\delta_{s's"}-i\epsilon_{kj}\sigma^3_{s's"}\right].\
\end{eqnarray}
\begin{eqnarray}
&&C(-q_1-q_2, s,s',q_1,i,q_2,j)-C(l-q_1-q_2, s,s',q_1,i,q_2,j)=g^2\frac{\sqrt{4\zeta_1\zeta_2}}{p^+}\frac{\zeta_1}{1-\zeta_1}\nonumber\\
&&\times\left[\frac{q_{1l}}{q_1^2}\frac{q_{2k}-(\zeta_1q_2-\zeta_2q_1)_k}{\zeta_2q_1^2  +\zeta_1q_2^2 -(\zeta_2q_1-\zeta_1q_2)^2}-\frac{\big(\zeta_1 l_l-q_{1l}\big)}{\left[(\zeta_1l-q_1)^2\right]}\frac{(\zeta_2l-q_2)_k+(\zeta_1q_2-\zeta_2q_1)_k}{\left[\zeta_2(\zeta_1l-q_1)^2  +\zeta_1(\zeta_2l-q_2)^2 -(\zeta_2q_1-\zeta_1q_2)^2\right]}\right]\nonumber\\
&&\times\left[\frac{2-\zeta_1}{\zeta_1}\delta_{li}\delta_{ss"}-i\epsilon_{li}\sigma^3_{ss"}\right]
\left[\frac{2(1-\zeta_1)-\zeta_2}{\zeta_2}\delta_{kj}\delta_{s''s'}-i\epsilon_{kj}\sigma^3_{s''s'}\right].\
\end{eqnarray}
Finally we can write
\begin{eqnarray}\label{f}
&&F(0,l,q_1,q_2,ij,s,s')=g^2\frac{\sqrt{4\zeta_1\zeta_2}}{p^+}\left[\frac{2-\zeta_1}{\zeta_1}\delta_{li}\delta_{ss"}-i\epsilon_{li}\sigma^3_{ss"}\right]
\left[\frac{2(1-\zeta_1)-\zeta_2}{\zeta_2}\delta_{kj}\delta_{s''s'}-i\epsilon_{kj}\sigma^3_{s''s'}\right]\nonumber\\
&&\Bigg\{\frac{\zeta_1}{1-\zeta_1}\left[\frac{q_{1l}}{q_1^2}\frac{q_{2k}-(\zeta_1q_2-\zeta_2q_1)_k}{\left[\zeta_2q_1^2  +\zeta_1q_2^2 -(\zeta_2q_1-\zeta_1q_2)^2\right]}-\frac{\big(\zeta_1 l_l-q_{1l}\big)}{\left[(\zeta_1l-q_1)^2\right]}\frac{(\zeta_2l-q_2)_k+(\zeta_1q_2-\zeta_2q_1)_k}{\left[\zeta_2(\zeta_1l-q_1)^2  +\zeta_1(\zeta_2l-q_2)^2 -(\zeta_2q_1-\zeta_1q_2)^2\right]}\right]\nonumber\\
&&+ \left[-\frac{q_{1l}}{q_1^2}-\frac{\zeta_1 l_l-q_{1l}}{(\zeta_1 l-q_1)^2}\right]\frac{(\zeta_2l-q_2)_k+(\zeta_1q_{2}-\zeta_2q_1)_k}{\Big((\zeta_2l-q_2)+(\zeta_1q_2-\zeta_2q_1)\Big)^2}\Bigg\}.\
\end{eqnarray}
Our goal now is to calculate the production cross section using \eq{sigma}. This involves symmetrizing the amplitude $F$ in \eq{f} and squaring it. 
The algebra is fairly lengthy, and we present the derivation in the Appendix.
The final result is 
 \begin{eqnarray} \label{c-2g}
 \frac{d\sigma^{qA\to \gamma(k_1)\gamma(k_2) X }}{d ^2{\bf k}_{1T} dk_1^+d^2{\bf k}_{2T} dk_2^+}&=&\frac{1}{2} \int d^{2}{\bf l}_T N({\bf l}_T)\Big[ F_{11}^2({\bf k}_{1T},\zeta_1;{\bf k}_{2T},\zeta_2; {\bf l}_T)+\bar F_{11}^2({\bf k}_{1T},\zeta_1;{\bf k}_{2T},\zeta_2; {\bf l}_T)+F_{12}^2({\bf k}_{1T},\zeta_1;{\bf k}_{2T},\zeta_2; {\bf l}_T) \nonumber\\
&&\,\,\,\,\,\,\,\,\,\,\,\,\,\,\,\,\,\,\,\,\,\,\,\,\,+\bar F_{12}^2({\bf k}_{1T},\zeta_1;{\bf k}_{2T},\zeta_2; {\bf l}_T)+({\bf k}_{1T},\zeta_1\leftrightarrow {\bf k}_{2T},\zeta_2)\Big], \
 \end{eqnarray}
where the functions $F_{\alpha\beta}$ and $\bar F_{\alpha\beta}$ are given in Eqs.\,(\ref{fn0},\ref{fn1},\ref{fn2},\ref{fn3})  in the Appendix. Note that in the final expersions for $F_{\alpha\beta}$ and $\bar F_{\alpha\beta}$  in Eqs.\,(\ref{fn0},\ref{fn1},\ref{fn2},\ref{fn3})  all  momenta $q, k_1, k_2, l$ are 2-dimensional vectors in the transverse plane. For clarity, in \eq{c-2g} and in the following, transverse momentum vector is denoted by a bold notation with a subscript $T$. 
 
In the next section we discuss our numerical analysis of \eq{c-2g}.  The simple case of the eikonal (soft) approximation where the analytical expersions can be significantly simplified,  is discussed in the Appendix.
 

\section{Numerical results and discussion}
The dipole amplitude $N({\bf l}_T)$ that enters \eq{c-2g} depends on energy through the longitudinal momentum variable $x_g(l_T)$ , which is defined as follows (see Appendix in \cite{di-photon}), 
\begin{eqnarray} \label{xqg}
x_g\equiv x_g\left(l_{T}; k_{1T}, \eta_{1}; k_{2T},\eta_{2}\right)&=&\frac{1}{x_q s}\Big[\frac{ k_{1T}^2}{\zeta_1}+ \frac{k_{2T}^2}{\zeta_2(1-\zeta_1)}  +\frac{|{\bf l}_T-{\bf k}_{1T}-{\bf k}_{2T}|^2}{1-\zeta_1-\zeta_2+\zeta_1 \zeta_2} \Big], \label{xg}  \nonumber\\
\zeta_1 &=&\frac{k_1^+}{p^+}= \frac{k_{1T}}{x_q\, \sqrt{s}}e^{\eta_{1}},\nonumber\\
\zeta_2 &=& \frac{k_2^+}{p^{+}-k_1^+}\approx \frac{k_2^+}{p^{+}}=\frac{k_{2T}}{x_q\,\sqrt{s}}e^{\eta_{2}}, \
\end{eqnarray} 
where the parameter $x_q$ is the ratio of the incoming quark to the projectile nucleon energy with $\sqrt{s}$ being the nucleon-nucleon center-of-mass energy. 
The parameter $x_g$ is the fraction of longitudinal momentum of the target carried by the gluon which is absorbed by the scattered quark. As discussed in the last paper in \cite{hybrid} this value of $x_g$ is appropriate if one assumes that all the momentum from the target is transferred to the quark in a single scattering event. Otherwise one should use a different value of $x_g$ consistent with multiple scattering off a dense target. However this issue is only important at NLO accuracy in $\alpha_s$, and in the present leading order calculation we follow the time honored practice and use the value given in \eq{xg}. The dependence of the dipole scattering amplitude on Bjorken $x_g$ is determined by the JIMWLK renormalization group equations.

The function $N\left(l_T, x_g\right)$ in \eq{c-2g} is related to the correlator $N(r,x_g)$ of two light-like
fundamental Wilson lines in the background of the color fields of the target nucleus (or proton)  through the Fourier transform. 
In the large $N_c$ limit, the coupled JIMWLK equations \cite{jimwlk} are simplified to the Balitsky-Kovchegov (BK) equation \cite{bk}, a closed-form equation for the rapidity evolution of the dipole amplitude which is presently known to next-to-leading accuracy \cite{Balitsky:2008zza,Kovner:2013ona}. The running-coupling improved BK equation (rcBK)  \cite{rcbk} has the same generic formal form as the leading-log BK evolution equation:
\begin{equation}
  \frac{\partial N(r,x)}{\partial\ln(x_0/x)}=\int d^2{\bf r_1}\
  K^{{\rm run}}({\bf r},{\bf r_1},{\bf r_2})
  \left[N(r_1,x)+N(r_2,x)
-N(r,x)-N(r_1,x)\,N(r_2,x)\right],
\label{bk1}
\end{equation}
where $r \equiv |\bf r_2-\bf r_1|$ is the transverse size of dipole and the $K^{{\rm run}}$ is evolution kernel  with Balitsky`s prescription \cite{bb} for the running coupling.  Note that inclusive di-photon production cross-section (both direct and fragmentation part)  depends on the dipole-target amplitude and therefore in principle probes the small-x dynamics. 
In the following we study the quantity, which is related to the above partonic cross section via
 \begin{equation}\label{photon-f3}
\frac{d\sigma^{pA\to \gamma(k_1)\gamma(k_2) X }}{d^2{\bf k}_{1T} d\eta_{1}d^2{\bf k}_{2T} d\eta_{2} }
= \int^1_{x_{q}^{min}} dx_q [f_q(x_q,\mu_I^2)+f_{\bar{q}}(x_{\bar{q}},\mu_I^2)] \, 
 \frac{d\sigma^{qA\to \gamma(k_1)\gamma(k_2) X }}{d^2{\bf k}_{1T} d\eta_{1}d^2{\bf k}_{2T} d\eta_{2} },
\end{equation}
where $f_q$ denotes the parton distributions function (PDF) in the projectile hadron (nucleon). The lower limit of integral  $x_{q}^{min}$ in the above is defined by, 
\begin{equation} \label{xqm}
x_q^{min} = Max\left( \frac{k_{1T} e^{\eta_{1}}}{\sqrt{s}}, \frac{k_{2T}e^{\eta_{2}}}{\sqrt{s}- k_{1T}e^{\eta_{1}}} \right). 
\end{equation}

Denoting the quantity in \eq{photon-f3} by the two photon differential cross section is a certain abuse of notation, since the actual cross section contains also contribution of photons emitted independently from two quarks (antiquarks). This extra contribution, although of the same order in $\alpha_{em}$ and $\alpha_s$ is of little interest to us, since we do not expect any correlated emission from independent quarks. It  will likely amount to an uncorrelated ``pedestal'' proportional to the product of single photon inclusive cross sections, which will have to be subtracted from the data (if and when such data is available) in order to compare with our results. Formally this subtraction can be written as 
\begin{equation}
\frac{d\sigma^{pA\to \gamma(k_1)\gamma(k_2) X }}{d^2{\bf k}_{1T} d\eta_{1}d^2{\bf k}_{2T} d\eta_{2} } \to \frac{d\sigma^{pA\to \gamma(k_1)\gamma(k_2) X }}{d^2{\bf k}_{1T} d\eta_{\gamma_1}d^2{\bf k}_{2T} d\eta_{2} }-\frac{1}{\sigma(|k_1|,|k_2|,\eta_{1},\eta_{2})}\frac{d\sigma^{pA\to \gamma(k_1) X }}{d^2{\bf k}_{1T} d\eta_{1}}\frac{d\sigma^{pA\to \gamma(k_2) X }}{d^2{\bf k}_{2T} d\eta_{2}}.
\end{equation}
The effective area $\sigma(|k_1|,|k_2|,\eta_{1},\eta_{2})$ is a slowly varying function of momenta (when the momenta are large enough) and is of the order of the area of the nucleus (or target proton). However its calculation from first principles requires an additional nonperturbative input, and rather than attempting it we suggest to treat it as an adjustable parameter in comparing with future data. In the rest of this paper we retain the notations of \eq{photon-f3}, while keeping the above caveat in mind.

The only external input for the rcBK non-linear evolution \eq{bk1} is the initial condition for the evolution which is taken to have the following form motivated by McLerran-Venugopalan (MV) model \cite{mv},  
  \begin{equation}
\mathcal{N}(r,Y\!=\!0)=
1-\exp\left[-\frac{\left(r^2\,Q_{0s}^2\right)^{\gamma}}{4}\,
  \ln\left(\frac{1}{\Lambda\,r}+e\right)\right],
\label{mv}
\end{equation}
where the onset of small-x evolution is assumed to be at $x_0=0.01$, and the infrared scale is taken $\Lambda=0.241$ GeV \cite{jav1}. 
The only free parameters in the above are $\gamma$ and the initial saturation scale $Q_{0s}$. 
The initial-condition of the rcBK equation has a non-perturbative nature and its parameters are fixed via a a global fit to proton structure functions in DIS in the small-x region \cite{jav1} and single inclusive hadron data in minimum-bias p+p collisions at RHIC and the LHC \cite{jav-d,pa-jam,pa-R}. Note that the current HERA data alone are not enough to uniquely fix the initial condition of the rcBK. Different parameter sets of the rcBK equation correspond to different initial saturation scale  $Q_{0s}$ (as probed by quarks). For proton-proton collisions, we take the initial saturation scale of proton $Q_{0s}^2\simeq 0.168\,\text{GeV}^2$  (with the corresponding $\gamma \simeq 1.119$) which was extracted from a global fit to proton structure functions in DIS in the small-x region \cite{jav1} and single inclusive hadron data in p+p collisions at RHIC and the LHC \cite{jav-d,pa-jam,pa-R}.   For the nucleus case, the initial saturation scale of a nucleus $Q_{0A}^2\approx 5\div 7 Q_{0s}^2$ should be considered as an impact-parameter averaged value and it is extracted from the minimum-bias data in deuteron-gold at RHIC and proton-lead collisions at the LHC \cite{pa-R}. In the absence of proper impact-parameter dependence of the dipole amplitude, rare events with higher multiplicity than the minimum-bias, may be described by using  a higher initial saturation scale incorporating somehow the effects of fluctuations and geometry \cite{jav-d,ridge6,hic-ap}, see also Refs.\,\cite{pp-LR, amir-hera1,amir-hera2,na-sat}.

\begin{figure}[t]                                                             
\includegraphics[width=8 cm] {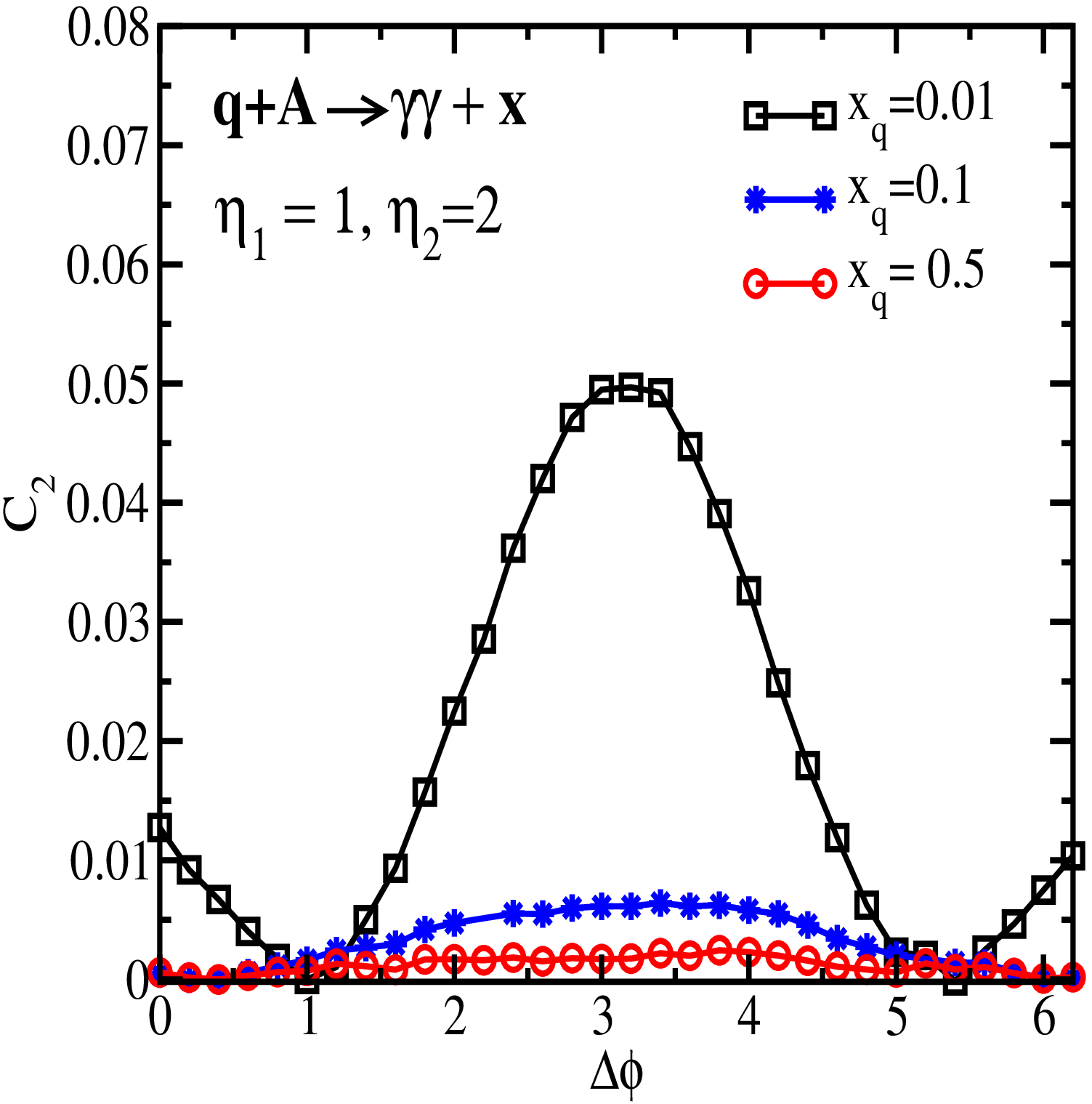} 
\includegraphics[width=8 cm] {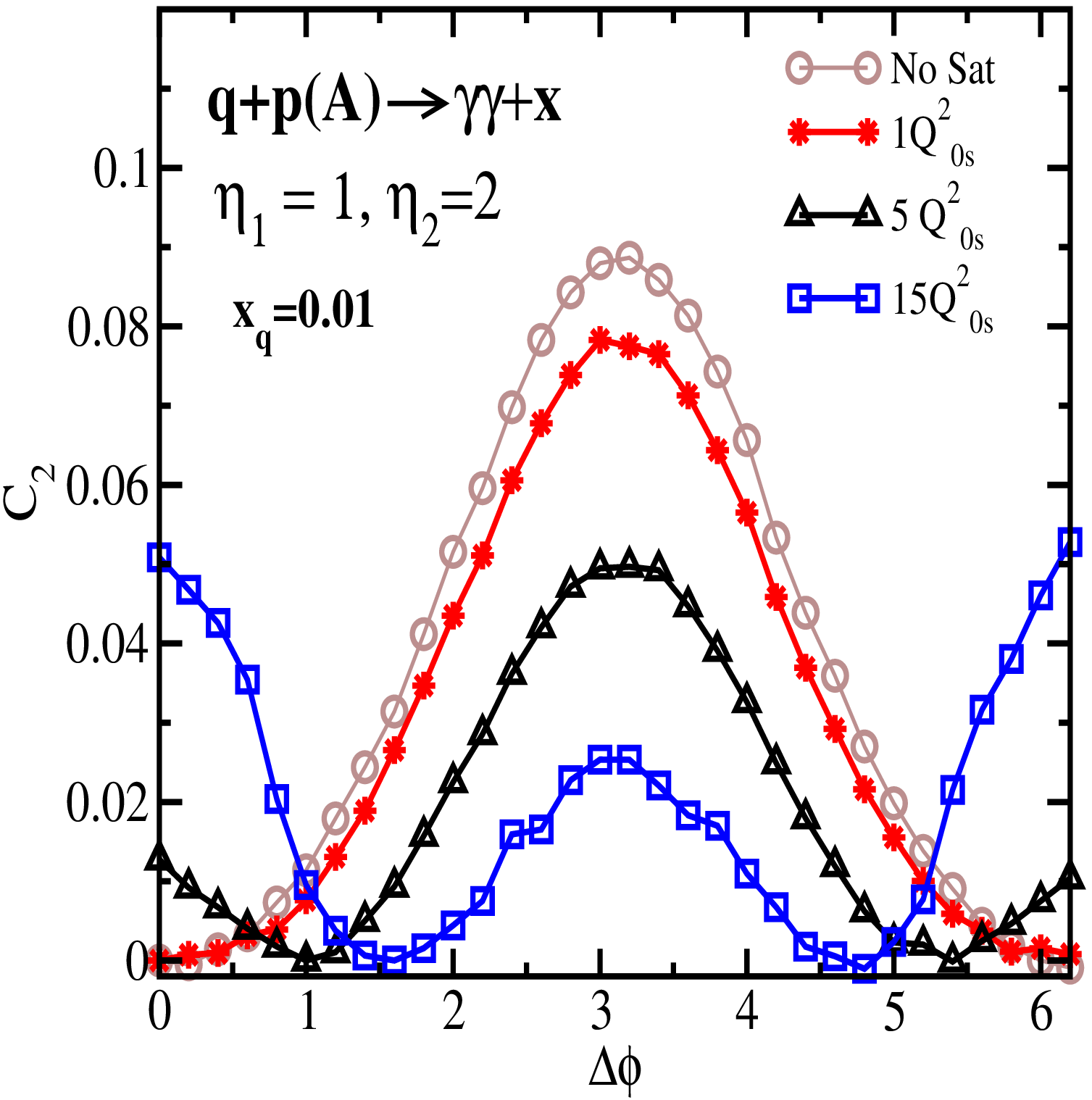}                             
\caption{ Left: the $x_q$ (ratio of energies of  the projectile quark to nucleon) dependence of prompt di-photon correlations $C_2$ obtained by solving the rcBK evolution equation with an initial saturation scale for the target  $Q^2_{0sA}= 5 Q^2_{0s}$ with $Q_{0s}^2=0.168~\text{GeV}^2$ corresponding to minimum-bias quark-nucleus collisions. Right: The initial-saturation-scale dependence of prompt di-photon correlations $C_2$ in q+p(A) collisions as a function of angle between two photons $\Delta\phi$. The results are obtained at a fixed $x_q=0.01$ by solving the rcBK evolution equation with different initial saturation scales for the target  $Q^2_{0sA}= N Q^2_{0s}$ with $N=1, 5,15$ and $Q_{0s}^2=0.168~\text{GeV}^2$. We also compare with the non-saturation (No Sat) model, see the text.   All curves in both panels are obtained at a fixed energy $\sqrt{s}=5.02$ TeV, at fixed rapidities of photons $\eta_1=1$  $\eta_2=2$ with transverse momenta of photons taken to be $k_{1T}=k_{2T}= 2$ GeV.  }
\label{f-qa1}
\end{figure}
Let us define the azimuthal correlation of the produced di-photon as \cite{amir-photon1,amir-photon2}, 
\begin{equation}\label{az}
C_2(\Delta \phi, k_{1T}, k_{2T},\eta_1,\eta_2)=
\frac{d\sigma^{pA\to \gamma(k_1)\gamma(k_2) X }}{d^2{\bf k}_{1T} d\eta_{\gamma_1}d^2{\bf k}_{2T} d\eta_{\gamma_2} }[\Delta \phi]/
\int_{0}^{2\pi} d\Delta \phi \frac{d\sigma^{pA\to \gamma(k_1)\gamma(k_2) X }}{  d^2{\bf k}_{1T} d\eta_{\gamma_1}d^2{\bf k}_{2T} d\eta_{\gamma_2} } -C_{ZYAM}, 
\end{equation}
where $\Delta \phi$  is the azimuthal angle between the two produced photons in the plane transverse to the collision axis. The azimuthal correlation $C_2$ is proportional to the relative probability of inclusive di-photon pair production in a given kinematics and angle  $\Delta \phi$ between the photons in the pair, normalized to the 
total cross-section integrated over angle $\Delta \phi$.  Following the experimental procedure for the zero-yield-at-minimum (ZYAM), we remove the constant background $C_{ZYAM}$ by shifting the  minimum at the zero axis. We expect that some of the theoretical uncertainties, such as sensitivity to possible $K$-factors which effectively incorporates the missing higher order corrections, drop out in the correlation defined in \eq{az}. One can equally take the normalization in \eq{az}  as the differential cross-section at a fixed reference angle  $\Delta \phi_c$ (away from near-side or away side). We checked that this  does not alter our over-all picture here.  One can equally define the correlation via coincidence probability \cite{di-e,ph-ex,amir-photon2}, however since the correlation defined via \eq{az} is  free from  extra integrals over transverse momenta, it should exhibit  the underlying dynamics of the correlation in a cleaner way. Note that defining the correlation for prompt di-photon via coincidence probability depends on the definition of the trigger single inclusive prompt photon whether it is taken to be direct or fragmentation prompt photon or include both contributions.  The correlation definition given in  \eq{az} is free from this ambiguity. 

\begin{figure}[t]                                       
                                \includegraphics[width=11 cm] {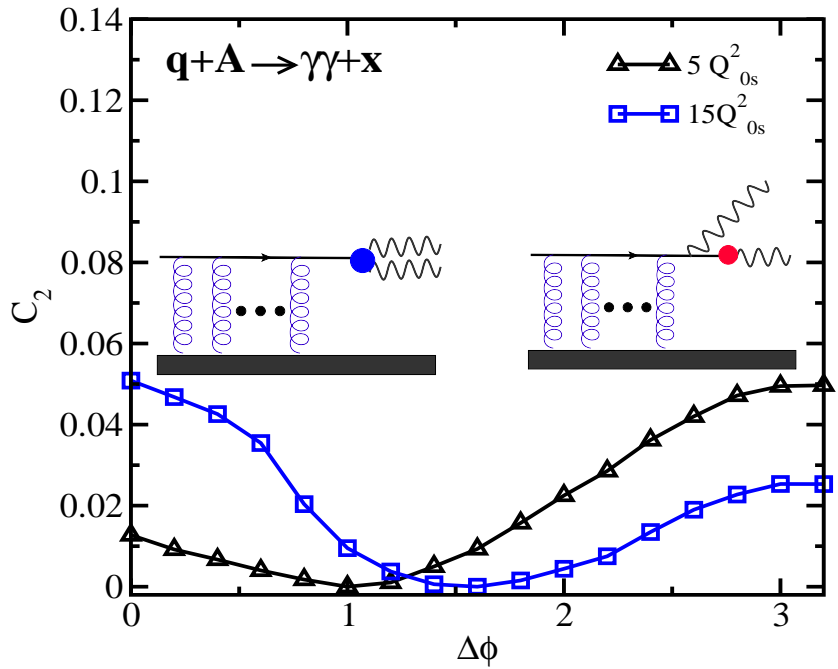}                                  
\caption{ Prompt di-photon  correlations $C_2$ in quark-nucleus collisions as a function of angle between two photons $\Delta\phi$ in q+Pb collisions at the LHC ($\sqrt{s}=5.02$ TeV) with $\eta_1=1, \eta_2=2$ and $x_q=0.01$. The curves are results of the rcBK solution with two initial saturation scale, similar to \fig{f-qa1} right panel. A schematic contribution of different typical CGC diagrams to the double inclusive prompt photon production at different angle is also shown.    }
\label{f-di}
\end{figure}

For consistency of our leading order approach, we should use the leading order PDF's. However the ratio \eq{az} depends on PDF's only very weekly.
 We checked that the leading-order (LO) and next-to-leading-order (NLO) PDFs give similar results for the di-photon cross-section, with less than $10\%$ discrepancy (which can be absorbed into the $K$-factor).  In the correlation $C_2$ defined via \eq{az}  the  effect of varying PDF's between LO and NLO is practically negligible within less than $2\%$ uncertainty. The results we present in the following were obtained using the NLO MSTW PDFs \cite{mstw}.  Following the conventional pQCD approach, we take the hard scale  $\mu_I = (k_{1T}+k_{2T})/2$ in the parton distribution in \eq{photon-f3}. For numerical computation, we focus here on low transverse momenta of the produced photon pair at the LHC at forward rapidities.  Note that this kinematics is mostly relevant for probing saturation effects and as we will demonstrate here, it is in fact the kinematics in which the ridge structure is by far the most pronounced.

We first focus on the prompt di-photon correlations in quark-nucleus (q+A) collisions. In \fig{f-qa1}, left panel, we show the di-photon correlations $C_2$ defined via \eq{az} in quark-nucleus collisions at the LHC energy $5.02$ TeV at fixed rapidities $\eta_1=1$ and $\eta_2=2$ for different values of $x_q$. We recall that the parameter $x_q$ is the ratio of the incoming quark to the projectile proton energy. All curves in \fig{f-qa1} left panel were obtained by solving the rcBK evolution equation with an initial saturation scale for the target  $Q^2_{0sA}= 5 Q^2_{0s}$ with $Q_{0s}^2=0.168~\text{GeV}^2$ corresponding to minimum-bias collisions.  It is seen that the near-side and away-side correlations (the peaks) is enhanced by decreasing $x_q$. For a large $x_q$, all correlations both at near-side and away-side disappear, albeit this effect is more pronounced for the near-side correlations.
Note that at fixed rapidity, energy and transverse momentum, a large $x_q$ corresponds to small $\zeta_{1}$ and $\zeta_{2}$, see \eq{xqg}. Our full numerical calculation for $C_2$ shown in \fig{f-qa1} is consistent with our theoretical conclusion about the absence of correlations at small $\zeta_i$ discussed in the Appendix.  

\begin{figure}[t]                                       
                                \includegraphics[width=9 cm] {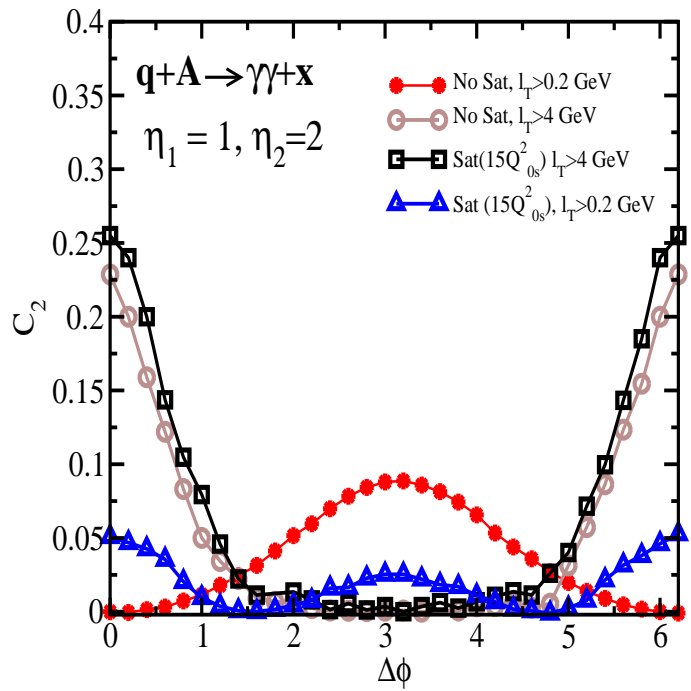}                               
\caption{The prompt di-photon correlations in q+A collisions as a function of angle between two photons $\Delta\phi$ obtained by imposing  different lower cut in the total transfer momentum to target  (or total transverse momentum of the produced quark-diphoton) denoted by $l_T^{min}$. 
The results are obtained within the rcBK saturation model with the initial saturation scale  $Q^2_{0sA}= 15 Q^2_{0s}$ and non-saturation model with leading-order perturbative unintegrated gluon density. All curves are obtained at $\sqrt{s}=5.02$ TeV with $x_q=0.01$, and $\eta_1=1$, $\eta_2=2$ and transverse momenta of photons $k_{1T}=k_{1T}= 2$ GeV.  }
\label{f-qa2}
\end{figure}

In \fig{f-qa1}, right panel, we show the initial saturation scale dependence of $C_2$ at a fixed 
$x_q=0.01$ in quark-proton and quark-nucleus  collisions at $5.02$ TeV with rapidities of the produced photon $\eta_1=1$ and  $\eta_2=2$. Different curves were obtained by  solving the rcBK evolution equation with different initial saturation scales for the target  $Q^2_{0sA}= N Q^2_{0s}$ with $N=1, 5,15$, and $Q_{0s}^2=0.168~\text{GeV}^2$ being the initial-saturation scale of the proton. We also compare with the non-saturation model (labeled by "No Sat" in the plot), by taking the unintegrated gluon distribution at the leading-order pQCD, $N(l_T,x_g)\propto 1/l_T^4$ in the prompt di-photon cross-section. It is seen that increasing the saturation scale in the system increases the near-side correlations leading to near-side collimation, while it decreases away-side correlation leading to the away-side decorrelation. 
Note that if momentum transfer from the target is small enough, the photon collinear to the outgoing quark, and the photon emerging from the initial photon-quark state have opposite transverse momenta, leading to back-to-back correlation at $\Delta\phi=\pi$ due to the single fragmentation di-photon contribution. Now increasing the exchanges to the target by increasing the saturation scale, unbalances the back-to-back correlations.  The double fragmentation and direct di-photon contributions (see Ref.\,\cite{di-photon}) on the other hand, are restricted to kinematics where the transverse momentum of the outgoing quark jet is relatively large. One therefore does not expect significant back-to-back correlation in the double fragmentation and direct di-photon contributions. By increasing the saturation scale,  the main contributions of integrand is shifted to higher transverse momentum, leading to the enhancement of double-fragmentation contributions at the near-side $\Delta\phi=0$.  The typical CGC diagrams which contribute at different $\Delta \phi$ are also shown in \fig{f-di}.  The curves in \fig{f-di} are the same as in \fig{f-qa1} right panel. 

In  \fig{f-qa2}  we show the effect of lower cut on the total momentum transfer to target $l_T$ in \eq{c-2g} (in the cross-section at partonic level) on the correlations $C_2$ within the saturation and non-saturation models. For the saturation model, we use the rcBK solution with initial saturation scale $Q^2_{0sA}= 15 Q^2_{0s}$ corresponding to a rare high-multiplicity event. All curves in \fig{f-qa2} are obtained at a fixed energy $\sqrt{s}=5.02$ TeV, at fixed rapidities of photons $\eta_1=1$ and $\eta_2=2$ with transverse momenta of photons taken to be $k_{1T}=k_{2T}= 2$ GeV. First it is seen that both saturation and non-saturation models produce similar correlations with a lower-cut $l_T>l_T^{min}=4$ GeV. This is clear since the saturation scale of the system in the chosen kinematics region is  smaller than $l_T^{min}$, and therefore saturation effects should not be important for $l_T>l_T^{min}$. Consequently both the saturation and the non-saturation models give similar results.   It is clearly seen in \fig{f-qa2}  that increasing  $l_T^{min}$ (from 0.2 GeV to 4 GeV) enhances the near-side correlations while decreases the  away-side correlations.  As we expect, the effect of changes on total momentum transfer to target (or changes on the lower-cut $l_T^{min}$) on the correlations is qualitatively similar to the effect of variation of the saturation scale $Q_s$ of the system.  

\begin{figure}[t]                                       
                                \includegraphics[width=8 cm] {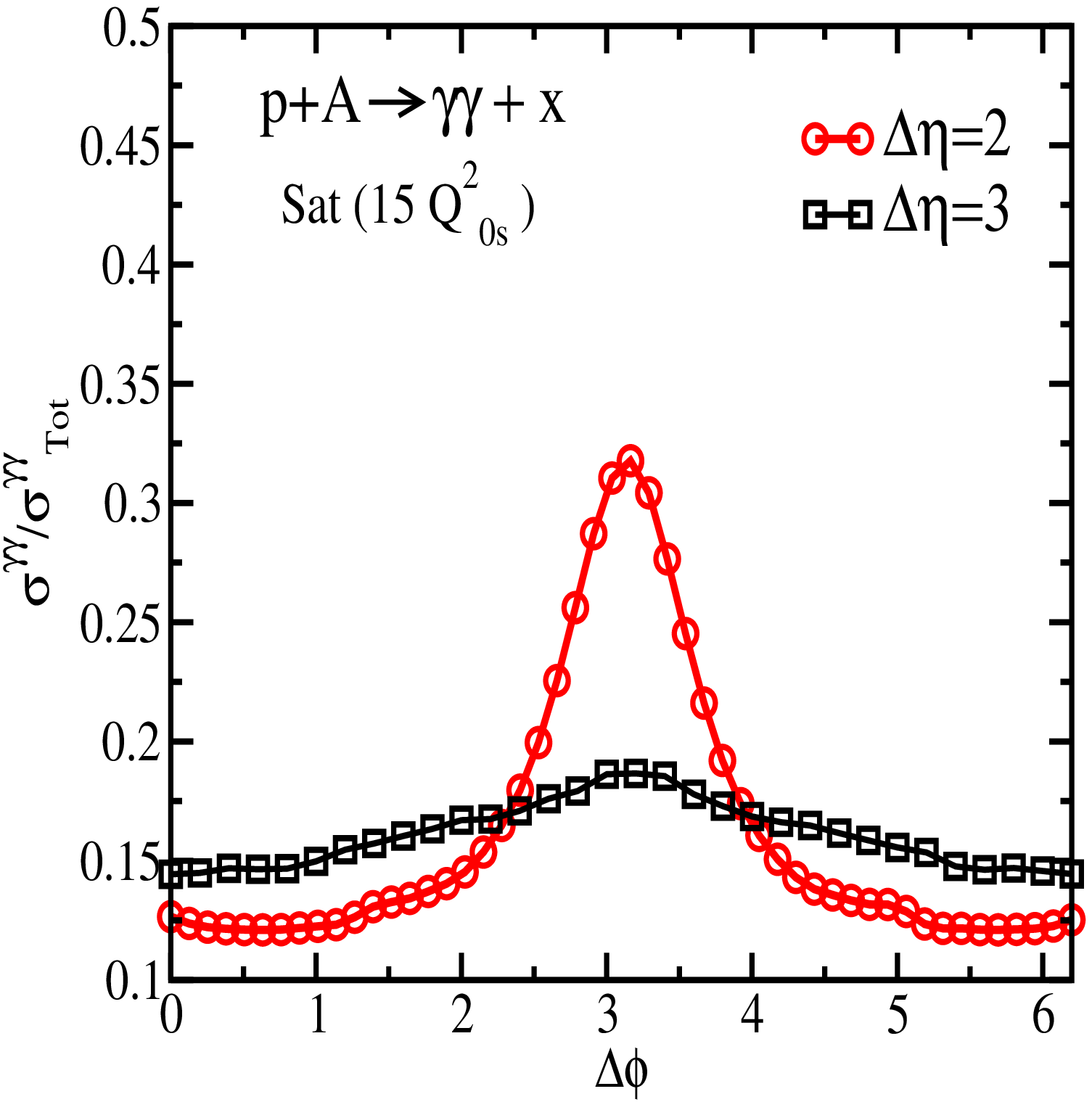} 
                                 \includegraphics[width=8 cm] {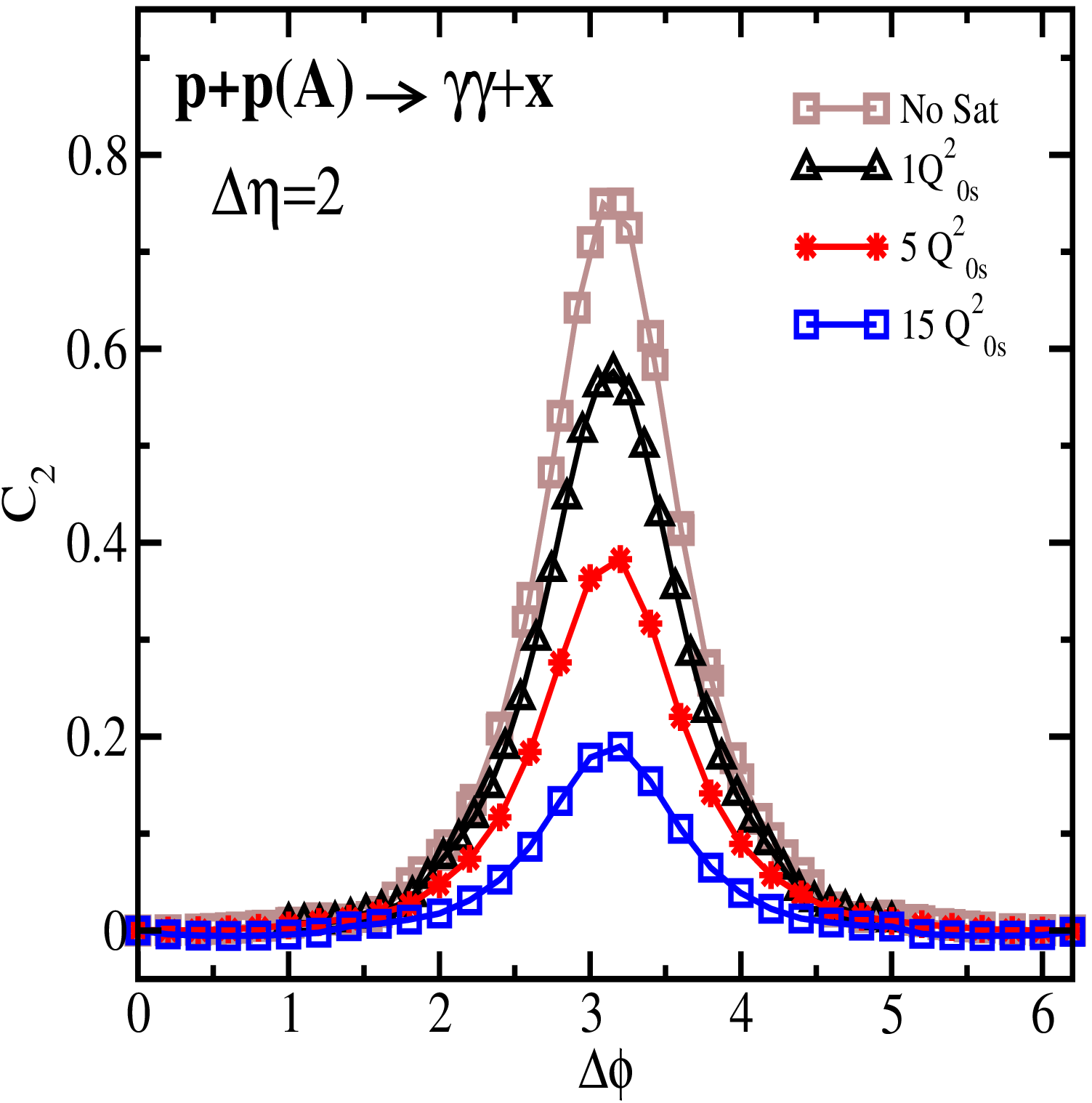}                                                
\caption{ Left: the rapidity dependence of the ratio of prompt di-photon differential cross-section in proton-nucleus collisions as a function of angle between two photons $\Delta\phi$, normalized to integrated cross-section over the angle. The curves are obtained for p+Pb collisions  by the rcBK solutions with the initial saturation scale $Q^2_{0sA}=15Q^2_{0s}$ Right: the initial-saturation-scale dependence of prompt di-photon correlations in p+p(A) collisions. The results are obtained by using the rcBK evolution equation solutions with different initial saturation scales for the target  $Q^2_{0sA}= N Q^2_{0s}$  and $N=1, 5,15$. We also compare  with the non-saturation (No Sat) model.   All curves are obtained at $\sqrt{s}=5.02$ TeV, and $\eta_1=1$, $\eta_2=\Delta \eta+\eta_1$ and transverse momenta of photons $k_{1T}=k_{1T}= 2$ GeV.}
\label{f-pa-all}
\end{figure}

Recall that as shown  above, the di-photon near-side collimation appears in quark-nucleus collisions only at small value of $x_q$. In order to compute the correlations in proton-nucleus collisions, we should convolute the  partonic cross-section  with the PDFs of quarks and anti-quarks, and perform the integral over $x_q$, see \eq{photon-f3}.  Note that although the partonic cross-section explicitly depends on $1/x_q^2$, in proton-nucleus collisions this will be modified due to $x_q$ dependence of the PDFs. Moreover,  the convolution integral in \eq{photon-f3} has a lower limit on $x_q$ via \eq{xqm} which depends on the energy/rapidity and transverse momentum of the produced photon.  Note also that the parameter $x_q$ also appears in the definition of $x_g$ via \eq{xqg} and hence the PDF convolution indirectly affects the rcBK evolution solutions of target nucleus in a non-trivial way (mainly due to energy-momentum conservation constraint). Therefore, a priori, it is not obvious how the integration over $x_q$ will affect the di-photon correlations in p+A collisions.

In  \fig{f-pa-all} left panel, we show the rapidity dependence of the ratio of prompt di-photon differential cross-section in proton-nucleus collisions as a function of angle between two photons $\Delta\phi$, normalized to integrated cross-section over the angle. The curves are obtained for p+A collisions at the LHC energy 5.02 TeV with fixed transverse momenta of photons $k_{1T}=k_{1T}= 2$ GeV. The initial saturation scale of the target was taken $Q^2_{0sA}=NQ^2_{0s}$ with $N=15$, corresponding to an event selection with high charged hadron multiplicity. Recall  that within the same multiplicity event selection (with $N=15$), the di-hadron ridge type structure is well pronounced \cite{ridge6}.  On the other hand we see here that within the same multiplicity event selection, the prompt di-photon near-side collimation almost disappears in p+A collisions at the LHC. In order to examine if the di-photon near-side correlations survive at any multiplicity event selections, in \fig{f-pa-all} right panel, we show the initial-saturation-scale dependence of prompt di-photon correlations defined via $C_2$ in p+p(A) collisions at 5.02 TeV and $\eta_1=1$, $\eta_2=3$ with fixed transverse momenta of photons $k_{1T}=k_{1T}= 2$ GeV.  The effect of size and density of the target is simulated by taking different initial saturation scale for the rcBK evolution equation.  Namely we use the different solutions of the rcBK evolution equation with different initial saturation scales for the target  $Q^2_{0sA}= N Q^2_{0s}$  and $N=1, 5,15$. We recall that an initial saturation scale $Q_{0s}$ describes a proton target in minimum-bias electron-proton or proton-proton collisions \cite{jav1}, while an initial saturation scale of about $5\div 7 Q_{0s}$ describes a heavy nuclear target in minimum-bias proton-nucleus collisions \cite{jav-d,pa-jam,pa-R}.  In order to examine the effect of saturation more clearly, we also show the results obtained by taking the dipole amplitude at the leading-order pQCD (labeled by "No Sat" in the plot). It is seen from \fig{f-pa-all}, while increasing  the saturation scale suppresses the away-side correlations, it does not affect the near-side correlation at the LHC energy 5.02 TeV in p+A collisions.    

\begin{figure}[t]                                       
         \includegraphics[width=8 cm] {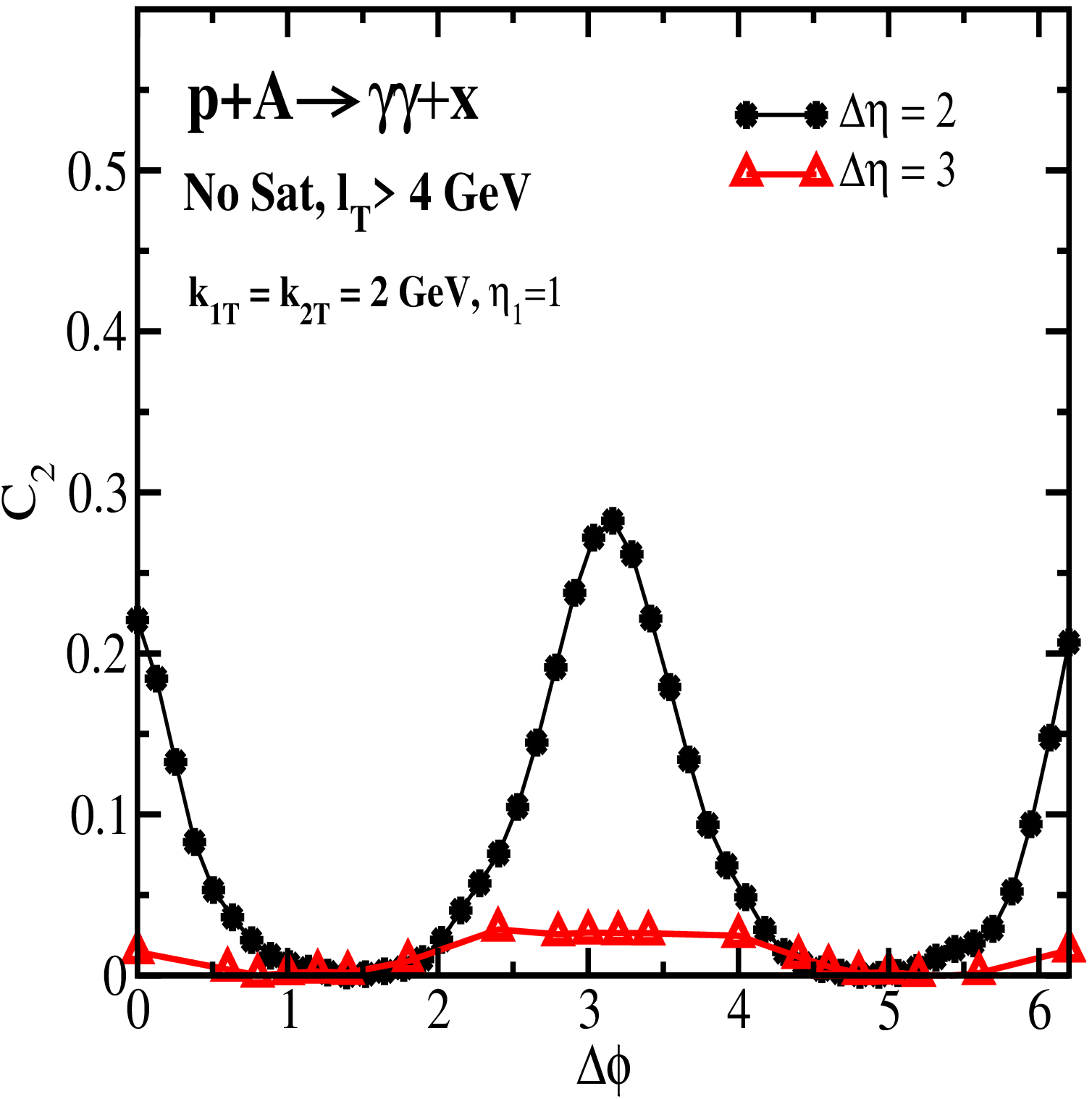}                         
\includegraphics[width=8 cm] {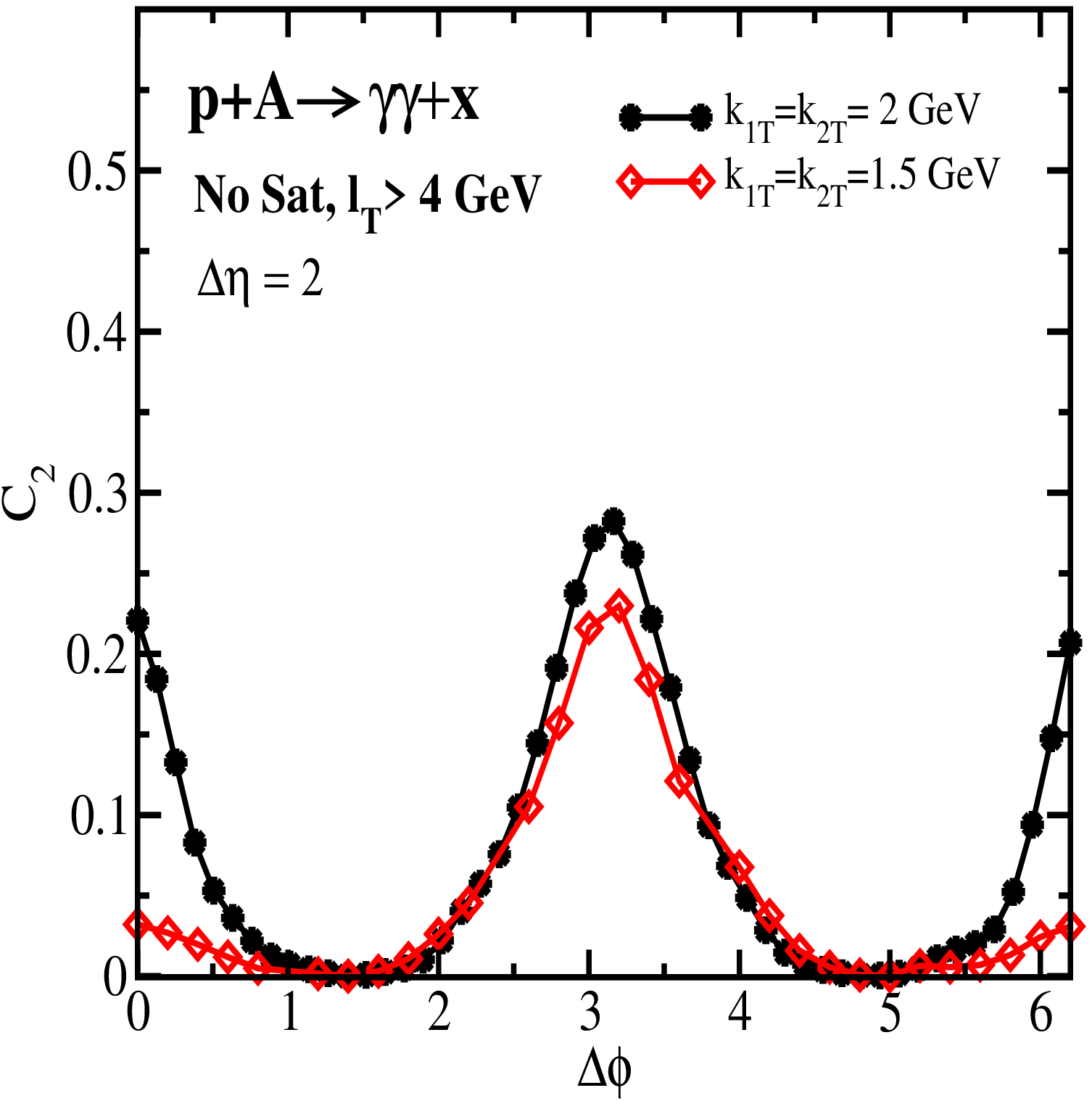} 
\caption{ The rapidity (left) and transverse momentum  (right) dependence of the prompt di-photon correlations in proton-nucleus collisions at 5.02 TeV as a function of angle between two photons $\Delta\phi$ obtained within the non-saturation model with a lower cutoff on the total transverse momentum of the produced quark-diphoton taken to be $l_T^{min}=4$ GeV.     }
\label{f-pa-cut}
\end{figure}

In order to examine the effect of a lower cut on the total transverse momentum $l_T^{min}$,  similar to calculations shown in \fig{f-qa2}, we imposed a lower cut $l_T^{min}=4$ GeV, and we computed the correlation  $C_2$ in the presence of such a cut in p+A collisions at the LHC energy 5.02 TeV.  In \fig{f-pa-cut}, we show the rapidity dependence (left panel) and transverse momenta of di-photon dependence (right panel) of the prompt di-photon correlation $C_2$ in proton-nucleus collisions within the non-saturation model with a lower cutoff.  It is again seen that similar to the case of quark-nucleus collisions, imposing a lower cut $l_T^{min}$, enhances the near-side correlations.  Note that in principle double photon fragmentation should not contribute when the rapidity of the produced photons are different\footnote{In principle, similar to Ref.\,\cite{di-photon}, one can explicitly extract from the prompt di-photon cross-section, single and double fragmentation functions and contributions, see Appendix B. However, this is not needed here as we stay away from the collinear kinematics.}. However, the remnant of the effect of double fragmentation may still persist up to a couple of units  in the rapidity difference (depending on the kinematics and energy). This is shown in \fig{f-pa-cut} left panel where we show $C_2$ as a function of $\Delta \eta$. It is seen that the correlations both in near-side and away-side in p+A collisions at the LHC disappears for approximately $\Delta \eta>3$. Therefore, these correlations are relatively short-range in nature and may not be considered as a genuine ridge compared to the di-hadron ridge. In principle, in order to enhance these correlations, one can choose a kinematics region more sensitive to the double fragmentation contribution. This can be achieved by triggering quarks (or hadrons) with a relative large transverse momentum simulating the effect of a large $l_T^{min}$. The exact value of $l_T^{min}$ depends on kinematics (and the acceptance of detector). The systematics of our results and the kinematics that the long-range correlations can be probed, will be outlined at the end of this section.

\begin{figure}[t]                                       
                                \includegraphics[width=8 cm] {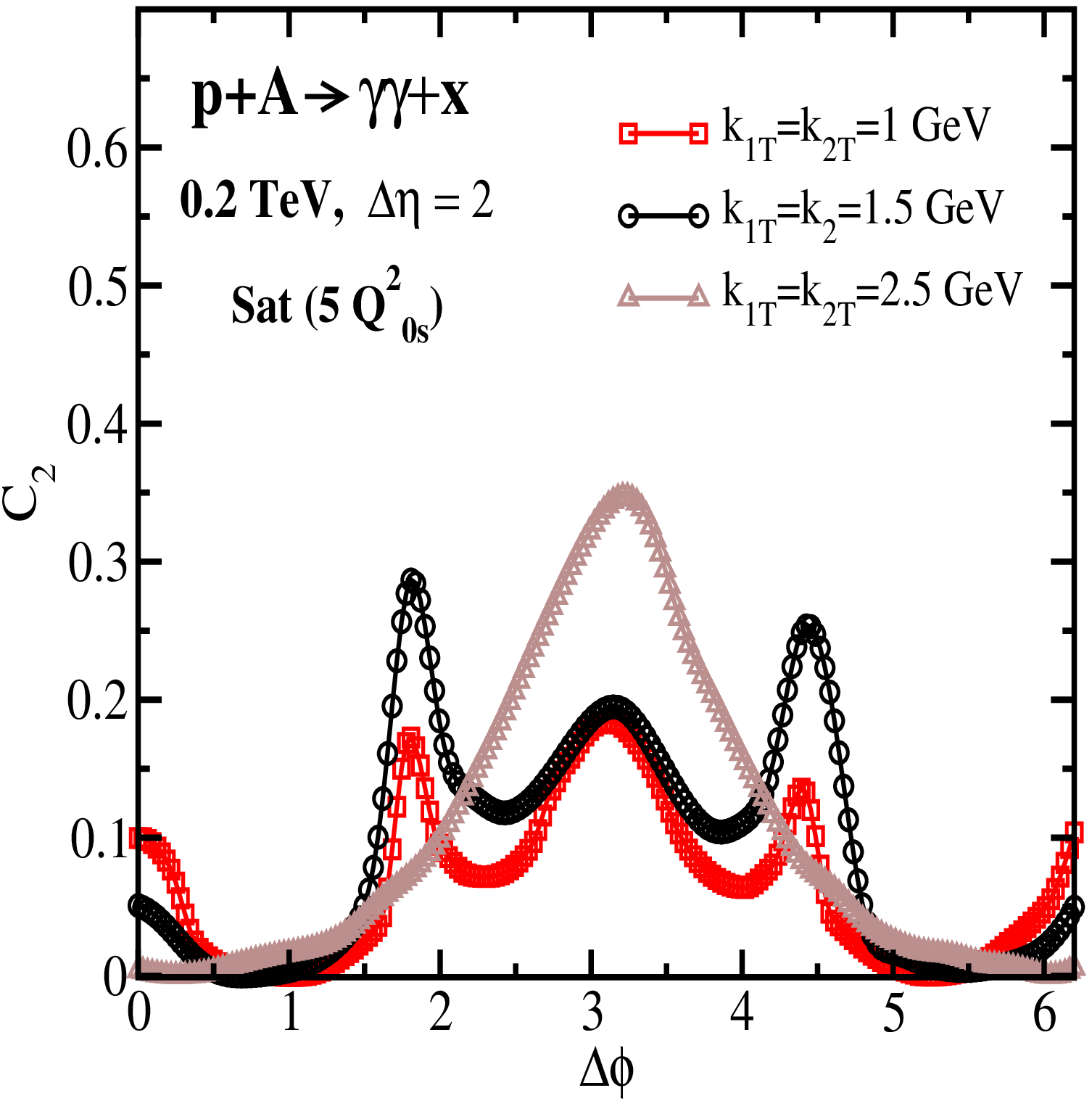}     
\includegraphics[width=8 cm] {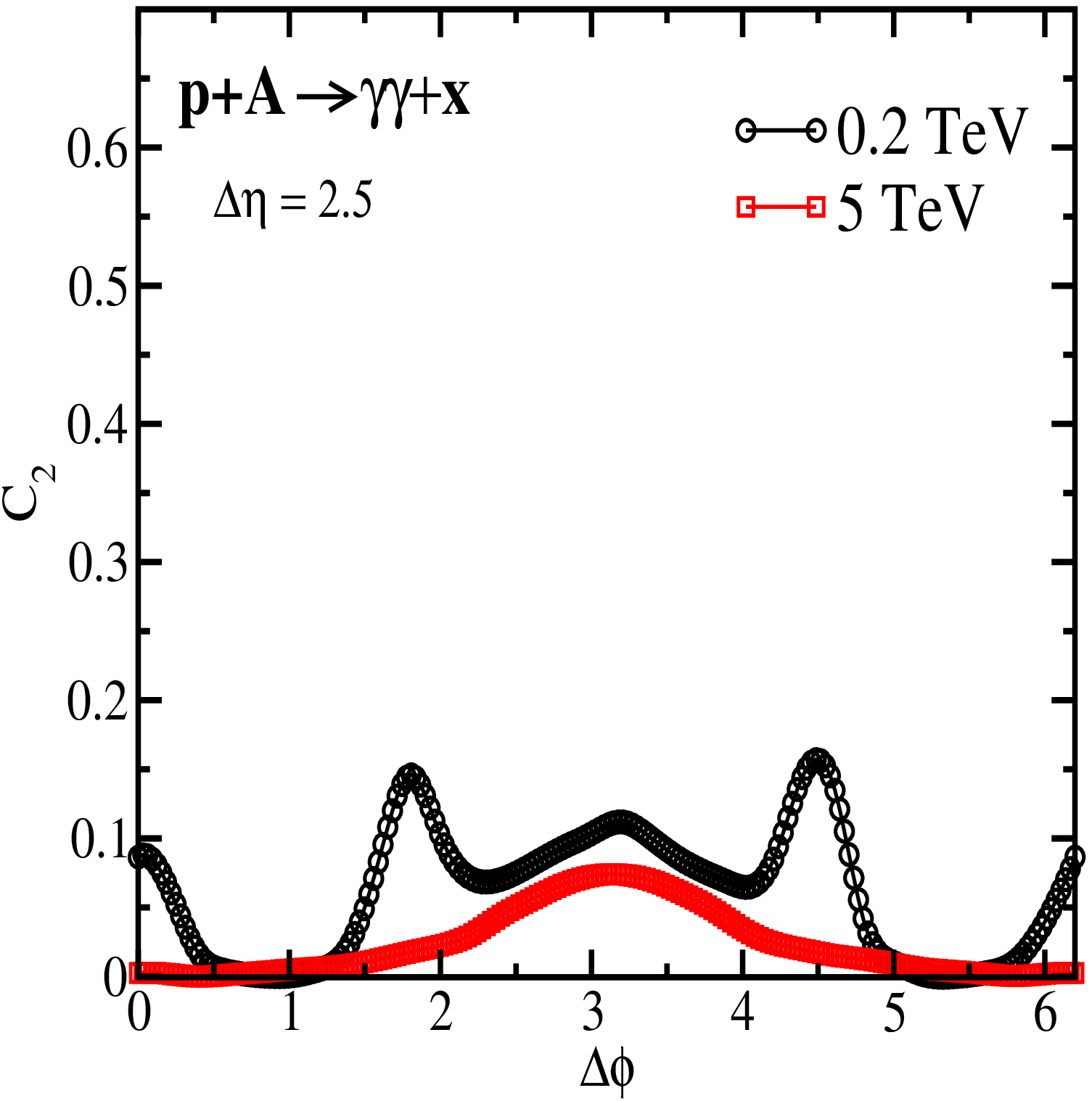}                                                                    
\caption{ Left: transverse momentum dependence of the prompt di-photon correlations $C_2$ in minimum-bias proton-nucleus collisions as a function of angle between two photons $\Delta\phi$ obtained with the rcBK solutions with the initial saturation scale $Q^2_{0sA}=5Q^2_{0s}$ at the RHIC energy 0.2 TeV with $\eta_1=1$ and $\eta_2=3$. Right:  comparison of $C_2$ at RHIC and the LHC for fixed  transverse momenta of di-photon at  $k_{1T}=k_{1T}= 1$ GeV with $\eta_1=1$ and $\eta_2=3.5$ in minimum-bias proton-nucleus collisions. }
\label{f-rhic}
\end{figure}

Note that the rapidity, energy and density  dependence of di-photon cross-sections and its evolution via the rcBK equation is very non-trivial due to complicated form of $x_g$ in \eq{xqg} that enters in the dipole amplitude in the cross-section. One also should bear in mind that generally, the small-x evolution in the rapidity at a fixed energy and transverse momentum is not the same as the evolution in energy but at a fixed rapidity.  This is partly due to the fact that the available phase space for particle production at different energy is different. 
Having this in mind, we next focus on prompt di-photon correlations at RHIC.  In \fig{f-rhic} right panel, we show the transverse momentum dependence of the prompt di-photon correlations $C_2$ in minimum-bias proton-nucleus collisions obtained with the rcBK solutions with the initial saturation scale $Q^2_{0sA}=5Q^2_{0s}$ (simulating minimum-bias p+A collisions with a heavy nuclear target) at the RHIC energy 0.2 TeV with $\eta_1=1$ and $\eta_2=3$.  In \fig{f-rhic} left panel, we compare the correlation $C_2$ at RHIC and the LHC at fixed  transverse momenta of di-photon $k_{1T}=k_{1T}= 1$ GeV  with rapidities $\eta_1=1$ and $\eta_2=3.5$ in minimum-bias proton-nucleus collisions. Note that all results in \fig{f-rhic} are obtained with imposing a lower cut  $l_T^{min}\sim \Lambda_{QCD}$ and the upper-cut  $l_T^{max}\sim 20$ GeV in the integral over $l_T$  at both RHIC and the LHC. It is interesting that at RHIC in stark contrast to the LHC, a ridge-like structure appears at low transverse momentum of the produced photons $0.2<k_T [\text{GeV}]<3$. The near-side collimation at RHIC survives up to rapidity interval $\Delta \eta\approx 3$.  
Another interesting features of the correlations at RHIC is the emergence of a double-peak structure for away-side correlations at low transverse momentum $k_T$ of photons. At higher $k_T$ it is replaced by a single peak structure due to the fragmentation contribution.  Note that a double-peak structure for the two-particle structure was also reported for the photon-hadron \cite{amir-photon1,amir-photon2} and dilepton-hadron \cite{dy-ana} correlations in a small windows of kinematics in the CGC approach. Therefore, it seems the existence of a double-peak structure at away-side to be a universal feature of electromagnetic probes. In the present case we understand its emergence in the following way. We recall that in order for higher Fock components of the projectile hadron wavefunction to be resolved and photons to be radiated, the projectile quark should interact with small-x target by exchanging transverse momentum. This means that when total momentum transfer $l_T$ is zero, the production rate of di-photon should go to zero and off-shell di-photon remains as a part of projectile hadron wavefunction.  Note however because of convolution with parton distribution functions, the local minimum at low $l_T$ will not be zero but gets smeared out.  On the other hand, the production has a maximum  at low $l_T$ due to single fragmentation contribution. Therefore, a double peak structure appears for the away side di-photon correlations as a result of combination of  a local minimum and a local maximum for the correlations. At higher transverse momentum, the back-to-back fragmentation contribution overwhelms the cross-section and prompt di-photon correlations has a single-peak structure.  Note that in the LHC kinematics, the typical momentum transfer to the target is larger than at RHIC and the double-peak structure becomes single-peak structure for a wide range of kinematics. This effect was also seen for the photon-hadron \cite{amir-photon1,amir-photon2} and dilepton-hadron \cite{dy-ana} correlations at RHIC and the LHC.

The systematics of the results can perhaps be understood in the following way. First, recall that in our approach the photons do not scatter off the hadronic target directly. The final state di-photons arise from two contributions. One is  the two photon component of the incoming quark state, which is put on shell by virtue of the quark scattering out of the initial state, while the other is the photon(s) emitted by the "bare" outgoing quark while it "reconstitutes" the dressed quark wave function, see \fig{f-di-cgc}. The first contribution is a part of the direct di-photon, while the second is mostly contained in the fragmentation (single and double fragmentation) part, see \fig{f-di}. There is no physics in the scattering event itself that can generate the same side correlations. Therefore such correlations can only appear due to correlations that exist in the proton wave function or are produced during the (double) fragmentation stage. 

 Just like in hadronic case Bose enhancement of photons exists  in the incoming quark (proton) wave function \cite{bose}. However now this enhancement does not extend over large interval of rapidities but is peaked at $\Delta\eta\sim 2$. Once the quark scatters out of the initial wave function putting  the two photons on shell, the initial state enhancement due to double fragmentation translates directly into  the same side correlations in the final state, since the momentum of the two photons is not changed by the scattering. In this sense the initial state correlations are even more directly reflected in the final state di-photon momentum distribution than in a di-hadron, which does absorb transverse momentum from the target. 

This above simple reasoning assumes that one can cleanly separate the final state photons which originate from the two photon initial state. This is difficult to do in the non-saturation model, the "No Sat" curves on Figs.\,(\ref{f-qa1},\ref{f-qa2}, \ref{f-pa-cut}) or when the total transfer momentum to target is very small $l_T^{min}\sim \Lambda_{QCD}$ (in the non-saturation model) in Figs.\,(\ref{f-qa2},\ref{f-pa-cut}). The scattering probability $N(l_T)\sim 1/l_T^4$ is strongly peaked at low momenta. On the other hand, if the momentum transfer to the quark is much smaller that the transverse momenta of the observed photons $k_T\ll k_{1T},k_{2T}$, while propagating through the target the quark changes direction adiabatically, so that no photons with momenta $k_{1T}\sim k_{2T}$ are produced.  The production is therefore dominated by the momentum transfer $k_T\sim k_{1T},\ k_{2T}$. In this regime however, there is no clear picture of a well defined contribution of the initial two photon state for two reasons. First, the factorization scale for fragmentation functions is itself of the order of the momentum transfer $\mu\sim k_T$ and with all momenta being of the same order some of the would be "initial state di-photons" are identified as fragmentation photons. Secondly, in this regime there is strong interference between different diagrams in the amplitude, and the separation on the probability level is not possible. As the saturation momentum is increased $Q_s>k_{1T},k_{2T}$ or $l_T^{min}>\Lambda_{QCD}$, the maximum of the scattering probability moves to higher transverse momentum transfer $k_T\sim Q_s$. At this point the simple picture of initial state correlations becomes valid, and one indeed observes the imprint of the initial correlations in the emitted di-photon distribution. Eventually, when $Q_s$ and therefore typical momentum transfer $k_T$ becomes too large, the quark is kicked out of the proton wave function too violently, so that an ever increasing contribution to the direct di-photon comes from the immediate vicinity of the interaction point rather than from the incoming wave function. Those photons are not correlated, and so the normalized correlation function $C_2$ decreases.

\section{Conclusion}
In this paper, we investigated prompt di-photon correlations, and in particular the existence of ridge correlations for di-photon production within the CGC approach in high-energy q+p(A) and p+p(A) collisions at RHIC and the LHC. Our formulation is only valid for the case of dilute-dense scatterings at forward rapidity.  Our main findings are as follows: 

  A ridge-type structure exists for prompt di-photon production in high-energy quark-nucleus collisions (even in minimum-bias collisions) at both the LHC and RHIC. Prompt di-photon ridge at near side $\Delta\phi\approx 0$ mainly comes from the double fragmentation di-photon contribution and the effects of Bose enhancement in the projectile wave function, while the away side ridge $\Delta\phi\approx \pi$ originates from the single fragmentation contribution which is dominated by the usual back-to-back correlations. The ridge structure at near-side  in quark-nucleus collisions disappears for rapidity differences $\Delta\eta\sim 3$ .  Additionally it exists only in a limited kinematics windows with small value of the light-cone variable $x_q$ and within the saturation regime, see Figs.\,(\ref{f-qa1},\ref{f-qa2}). 

For the case of proton-nucleus collision, the near-side correlations are washed away at the LHC 5.02 TeV after integration over $x_q$ variable (or convolution with the PDFs) while the ridge-like structure survives in p+A collisions at RHIC 0.2 TeV at rather low transverse momenta of the produced photons even up to $\Delta \eta \approx 2\div 3$,
 see Figs.\,(\ref{f-pa-all},\ref{f-rhic}). We showed that in principle the ridge-like structure at the LHC can re-appear by imposing a lower cut on the total transfer momentum to the target $l_T^{min}$ by isolation cut techniques, see \fig{f-pa-cut}.   

The prompt di-photon near side $\Delta\phi\approx 0$ correlations (the ridge) strongly depends on the saturation dynamics and the value of typical total transfer momentum to the target (or $l_T^{min}$). It is absent when $Q_s\ll k_T$, is generated by increasing the saturation scale to $k_T< Q_s$  or $k_T<l_T^{min}$ (at small $x_q$). This is mainly due to the fact that a larger saturation scale shifts the main contribution of the integrand of the cross-section to higher transverse momentum where the double-fragmentation becomes important leading to enhancement of near-side correlations while it reduces the single fragmentation contribution, consequently unbalancing the back-to-back correlations leading to suppressions of the way-side correlations, see e.g. \fig{f-di}.
 
The prompt di-photon correlations at forward rapidities in the away side region  exhibit single or double peak structure depending on the kinematics, namely transverse momenta of the produced photons, rapidity interval between two photons and center-of-mass energy. Similar effect was also reported for photon-hadron \cite{amir-photon1,amir-photon2} and dilepton-hadron \cite{dy-ana} correlations in high-energy p+Pb collisions.

Although di-hadron ridge was observed  in p+p and p(d)+A collisions \cite{exp-pp,exp-pa1,exp-pa2,exp-pa3,exp-pa4,exp-pa5,exp-pa6} both at the LHC and RHIC, the di-photon correlations have not yet been measured at the LHC and RHIC.  Our study here shows that  the energy, density and transverse momenta dependence of di-photon correlations both at near side and away side provide essential complementary information to understand the true underlying dynamics of emergence of ridge phenomenon in high-energy collisions.  It was argued that the di-hadron ridge effect can be understood entirely within the framework of the initial-state CGC physics. There are also competing mechanisms based on the final state effects, like Hydrodynamics approach \cite{hydro-1,hydro-2}, which provide excellent description of the same data. Our study here pertains to the minimal bias p-Pb data, where one certainly does not expect collective phenomena in the final state to play a crucial role. Therefore observation or non observation of di photon correlations at RHIC and the LHC (see e.g. \fig{f-rhic}) would be an important test for the two approaches.

It is interesting to note that  Ref.\,\cite{ridge6} required both the projectile proton and the target to be in the saturation regime in order to describe the ridge effect in p+Pb collisions.  The existence of two saturation scales in the problem (for the projectile proton and the target)  both in p+p and p+Pb collisions, is essential  for \,\cite{ridge6} in order to find a good fit to the observed number of charged hadron tracks in which the ridge appears. The origin of such rare high multiplicity events in p+p and p+Pb collisions is not yet well understood. In our hybrid approach on the other hand the projectile proton is assumed to be in dilute regime and is consequently treated in standard parton model approach, while the target is treated as a dense object. Therefore, we have here only one saturation scale and our calculation should be applicable to minimal bias events in p+Pb collisions and a more significant fraction of p+p events than those exhibiting di-hadron ridge correlations. In this sense the di-photon correlations is a much cleaner probe of initial state effects compared to di-hadrons, as it does not require modelling of rare fluctuations leading to high multiplicity events.

One of the main results of  this paper is that the prompt di-photon near-side collimation and the ridge-type structure can also be probed in quark-nucleus collisions at the LHC at small value of $x_q$, albeit  within a limited rapidity interval. We recall that strictly speaking  our formulation is only valid for dilute-dense scatterings (such as proton-nucleus and electron-nucleus collisions) at small-x. Therefore, our general outcome here in particular the di-photon ridge phenomenon in quark-nucleus collisions can be also tested in a future electron-ion collider \cite{eic1,eic2}. A detailed study of di-photon correlations in high-energy electron-nucleus collisions is postponed in future publication. 

\begin{acknowledgments}
The work of A.K. is supported by the DOE grant DE-FG02-13ER41989. The work of A.H.R. is supported in part by Fondecyt grant 1110781, 1150135 and Conicyt C14E01.
\end{acknowledgments}

\newpage
\section*{APPENDIX}
In this Appendix we present the details of the calculation of the di-photon production cross-section following the formalism introduced in Sec. II.
Symmetrizing \eq{f} we obtain:

\begin{eqnarray}\label{fsim}
&&F(0,l,q_1,q_2,ij,s,s')+F(0,l,q_2,q_1,ji,s,s')=g^2\frac{\sqrt{4\zeta_1\zeta_2}}{p^+}\left[\frac{2-\zeta_1}{\zeta_1}\delta_{li}\delta_{ss"}-i\epsilon_{li}\sigma^3_{ss"}\right]
\left[\frac{2(1-\zeta_1)-\zeta_2}{\zeta_2}\delta_{kj}\delta_{s''s'}-i\epsilon_{kj}\sigma^3_{s''s'}\right]\nonumber\\
&&\times\Bigg\{\frac{\zeta_1}{1-\zeta_1}\left[\frac{q_{1l}}{q_1^2}\frac{q_{2k}-(\zeta_1q_2-\zeta_2q_1)_k}{\left[\zeta_2q_1^2  +\zeta_1q_2^2 -(\zeta_2q_1-\zeta_1q_2)^2\right]}-\frac{\big(\zeta_1 l_l-q_{1l}\big)}{\left[(\zeta_1l-q_1)^2\right]}\frac{(\zeta_2l-q_2)_k+(\zeta_1q_2-\zeta_2q_1)_k}{\left[\zeta_2(\zeta_1l-q_1)^2  +\zeta_1(\zeta_2l-q_2)^2 -(\zeta_2q_1-\zeta_1q_2)^2\right]}\right]\nonumber\\
&&+ \left[-\frac{q_{1l}}{q_1^2}-\frac{\zeta_1 l_l-q_{1l}}{(\zeta_1 l-q_1)^2}\right]\frac{(\zeta_2l-q_2)_k+(\zeta_1q_{2}-\zeta_2q_1)_k}{\Big((\zeta_2l-q_2)+(\zeta_1q_2-\zeta_2q_1)\Big)^2}\Bigg\}\nonumber\\
&&+\frac{\sqrt{4\zeta_1\zeta_2}}{p^+}\left[\frac{2(1-\zeta_2)-\zeta_1}{\zeta_1}\delta_{ki}\delta_{s''s'}-i\epsilon_{ki}\sigma^3_{s''s'}\right]\left[\frac{2-\zeta_2}{\zeta_2}\delta_{lj}\delta_{ss"}-i\epsilon_{lj}\sigma^3_{ss"}\right]\nonumber\\
&&\times\Bigg\{\frac{\zeta_2}{1-\zeta_2}\left[\frac{q_{1k}-(\zeta_2q_1-\zeta_1q_2)_k}{\left[\zeta_1q_2^2  +\zeta_2q_1^2 -(\zeta_1q_2-\zeta_2q_1)^2\right]}\frac{q_{2l}}{q_2^2}-\frac{(\zeta_1l-q_1)_k+(\zeta_2q_1-\zeta_1q_2)_k}{\left[\zeta_1(\zeta_2l-q_2)^2  +\zeta_2(\zeta_1l-q_1)^2 -(\zeta_1q_2-\zeta_2q_1)^2\right]}\frac{\big(\zeta_2 l_l-q_{2l}\big)}{\left[(\zeta_2l-q_2)^2\right]}\right]\nonumber\\
&&+ \frac{(\zeta_1l-q_1)_k+(\zeta_2q_{1}-\zeta_1q_2)_k}{\Big((\zeta_1l-q_1)+(\zeta_2q_1-\zeta_1q_2)\Big)^2}\left[-\frac{q_{2l}}{q_2^2}-\frac{\zeta_2 l_l-q_{2l}}{(\zeta_2 l-q_2)^2}\right]\Bigg\}.\
\end{eqnarray}
We will calculate the square of the first three lines in \eq{fsim} and the product of the first three lines  by the last three lines. The complete result will then be given by adding the symmetrizing term.
In the first part of the calculation we encounter the expressions of the type
\begin{eqnarray}
X_1&\equiv&\frac{1}{2}tr\left\{\left[\frac{(2-\zeta_1)(2(1-\zeta_1)-\zeta_2)}{\zeta_1\zeta_2}a_ib_j-\epsilon_{ik}a_k\epsilon_{jl}b_l\right]\delta+i\left[\frac{2-\zeta_1}{\zeta_1}a_i\epsilon_{jk}b_k+\frac{2(1-\zeta_1)-\zeta_2}{\zeta_2}\epsilon_{il}a_lb_j\right]\sigma^3\right\}\nonumber\\
&&\times\left\{\left[\frac{(2-\zeta_1)(2(1-\zeta_1)-\zeta_2)}{\zeta_1\zeta_2}c_id_j-\epsilon_{ik}c_k\epsilon_{jl}d_l\right]\delta-i\left[\frac{2-\zeta_1}{\zeta_1}c_i\epsilon_{jk}d_k+\frac{2(1-\zeta_1)-\zeta_2}{\zeta_2}\epsilon_{il}c_ld_j\right]\sigma^3\right\},\\
X_2&\equiv&\frac{1}{2}tr\left\{\left[\frac{(2-\zeta_1)(2(1-\zeta_1)-\zeta_2)}{\zeta_1\zeta_2}a_ib_j-\epsilon_{ik}a_k\epsilon_{jl}b_l\right]\delta+i\left[\frac{2-\zeta_1}{\zeta_1}a_i\epsilon_{jk}b_k+\frac{2(1-\zeta_1)-\zeta_2}{\zeta_2}\epsilon_{il}a_lb_j\right]\sigma^3\right\}\nonumber\\
&&\times\left\{\left[\frac{(2-\zeta_2)(2(1-\zeta_2)-\zeta_1)}{\zeta_2\zeta_1}c_id_j-\epsilon_{ik}c_k\epsilon_{jl}d_l\right]\delta-i\left[\frac{2(1-\zeta_2)-\zeta_1}{\zeta_1}c_i\epsilon_{jk}d_k+\frac{2-\zeta_2}{\zeta_2}\epsilon_{il}c_ld_j\right]\sigma^3\right\},\
\end{eqnarray}
 where $a,b,c,d$ are various transverse momenta. After some tedious algebra this can be explicitly written as 
 \begin{eqnarray}
 X_1&=&\Sigma(12;12)(a\cdot c)(b\cdot d)-\bar\Sigma(12;12)(a\times c)(b\times d),\nonumber\\
 X_2&=& \Sigma(12;21)(a\cdot c)(b\cdot d)-\bar\Sigma(12;21)(a\times c)(b\times d).\
 \end{eqnarray}
 Note that the second part of the above expressions can be further simplified by using the following identity,  
 \begin{equation} (a\times c)(b\times d)=(a\cdot b)(c\cdot d)-(a\cdot d)(b\cdot c).\end{equation}
 Let us introduce the following notation
 \begin{eqnarray}
&& \Sigma(12;12)=\frac{\left[(2-\zeta_1)^2+\zeta_1^2\right]\left[(2-2\zeta_1-\zeta_2)^2+\zeta_2^2\right]}{\zeta_1^2\zeta_2^2},\nonumber\\
&&\bar\Sigma(12;12)=\frac{4(2-\zeta_1)(2-2\zeta_1-\zeta_2)}{\zeta_1\zeta_2},\nonumber\\
&&\Sigma(12;21)=\frac{\left[(2-\zeta_1)(2(1-\zeta_2)-\zeta_1)+\zeta_1^2\right]\left[(2-\zeta_2)(2(1-\zeta_1)-\zeta_2)+\zeta_1^2\right]}{\zeta_1^2\zeta_2^2},\nonumber\\ 
&&\bar\Sigma(12;21)=\frac{4(2-\zeta_1-\zeta_2)^2}{\zeta_1\zeta_2}.\
\end{eqnarray}
 We multiply out the different terms
 \begin{eqnarray}  \label{fn0}
 &&F^2_{11}=g^4\frac{4\zeta_1\zeta_2}{p^{+2}}\Sigma(12;12)\Bigg[\nonumber\\
 &&\frac{\zeta_1^2}{(1-\zeta_1)^2}\Bigg(\frac{\left[q_2-(\zeta_1q_2-\zeta_2q_1)\right]^2}{q_1^2\left[\zeta_2q_1^2  +\zeta_1q_2^2 -(\zeta_2q_1-\zeta_1q_2)^2\right]^2}+\frac{[(\zeta_2l-q_2)+(\zeta_1q_2-\zeta_2q_1)]^2}{\left[(\zeta_1l-q_1)^2\right]\left[\zeta_2(\zeta_1l-q_1)^2  +\zeta_1(\zeta_2l-q_2)^2 -(\zeta_2q_1-\zeta_1q_2)^2\right]^2}\nonumber\\
 &&- \frac{2\big[q_1\cdot\big(\zeta_1 l-q_1\big)\big]\big[\big(q_{2}-(\zeta_1q_2-\zeta_2q_1)\big)\cdot\big((\zeta_2l-q_2)+(\zeta_1q_2-\zeta_2q_1)\big)\big]}
 {q_1^2\left[\zeta_2q_1^2  +\zeta_1q_2^2 -(\zeta_2q_1-\zeta_1q_2)^2\right]\left[(\zeta_1l-q_1)^2\right]\left[\zeta_2(\zeta_1l-q_1)^2  +\zeta_1(\zeta_2l-q_2)^2 -(\zeta_2q_1-\zeta_1q_2)^2\right]} \Bigg)\nonumber\\
 &&+\frac{\zeta_1^2l^2}{q_1^2(\zeta_1l-q_1)^2\Big((\zeta_2l-q_2)+(\zeta_1q_2-\zeta_2q_1)\Big)^2}\nonumber\\
 &&+\frac{2\zeta_1}{1-\zeta_1}\Bigg(\frac{\Big[\zeta_1l\cdot(\zeta_1l-q_1)\Big]\Big[((\zeta_1q_2-\zeta_2q_1)-q_{2})\cdot ((\zeta_2l-q_2)+(\zeta_1q_{2}-\zeta_2q_1))\Big]}{q_1^2[\zeta_1l-q_1]^2\left[\zeta_2q_1^2  +\zeta_1q_2^2 -(\zeta_2q_1-\zeta_1q_2)^2\right]\Big((\zeta_2l-q_2)+(\zeta_1q_2-\zeta_2q_1)\Big)^2}\nonumber\\
 &&+
 \frac{\zeta_1(l\cdot q_1)}{q_1^2[\zeta_1l-q_1]^2\left[\zeta_2(\zeta_1l-q_1)^2  +\zeta_1(\zeta_2l-q_2)^2 -(\zeta_2q_1-\zeta_1q_2)^2\right]}\Bigg)\Bigg].\
 \end{eqnarray}
 \begin{eqnarray}  \label{fn1}
 &&\bar F^2_{11}=-g^4\frac{8\zeta_1\zeta_2}{p^{+2}}\bar\Sigma(12;12)\Bigg[\nonumber\\
 &&-\frac{\zeta_1^2}{(1-\zeta_1)^2}\frac {\zeta_1\zeta_2\Big[( q_1\times l)\Big]\Big[(1-\zeta_1)(q_2\times l)+\zeta_2(q_1\times l)\Big] }{q_1^2\left[(\zeta_1l-q_1)^2\right]\left[\zeta_2q_1^2  +\zeta_1q_2^2 -(\zeta_2q_1-\zeta_1q_2)^2\right]\left[\zeta_2(\zeta_1l-q_1)^2  +\zeta_1(\zeta_2l-q_2)^2 -(\zeta_2q_1-\zeta_1q_2)^2\right]}
 \nonumber\\
 &&-\frac{\zeta_1}{(1-\zeta_1)}\frac{\zeta_1\zeta_2\Big[( q_1\times l)\Big]\Big[(1-\zeta_1)(q_{2}\times l)+\zeta_2(q_1\times l)\Big]}{q_1^2\Big[(\zeta_1 l-q_1)^2\Big]\left[\zeta_2q_1^2  +\zeta_1q_2^2 -(\zeta_2q_1-\zeta_1q_2)^2\right]\Big((\zeta_2l-q_2)+(\zeta_1q_2-\zeta_2q_1)\Big)^2}\Bigg]\nonumber\\
 &&= \frac{8\zeta_1^3\zeta_2^2}{(1-\zeta_1)p^{+2}}\bar\Sigma(12;12)\frac{\Big[(1-\zeta_1)[q_1\cdot q_2 l^2 -(q_1\cdot l)(q_2\cdot l)] +\zeta_2[q_1^2l^2-(q_1\cdot l)^2]\Big]}{q_1^2\Big[(\zeta_1 l-q_1)^2\Big]\left[\zeta_2q_1^2  +\zeta_1q_2^2 -(\zeta_2q_1-\zeta_1q_2)^2\right]}\Bigg]\nonumber\\
 &&\times \left[\frac{\zeta_1}{1-\zeta_1}\frac{1}{\left[\zeta_2(\zeta_1l-q_1)^2  +\zeta_1(\zeta_2l-q_2)^2 -(\zeta_2q_1-\zeta_1q_2)^2\right]}+\frac{1}{\Big((\zeta_2l-q_2)+(\zeta_1q_2-\zeta_2q_1)\Big)^2}\right].\
 \end{eqnarray}
 \begin{eqnarray} \label{fn2}
  &&F^2_{12}=g^4\frac{4\zeta_1\zeta_2}{p^{+2}}\Sigma(12;21)\Bigg[\frac{\zeta_1\zeta_2}{(1-\zeta_1)(1-\zeta_2)}\Bigg\{\frac{[(1-\zeta_2)q_1^2+\zeta_1q_1\cdot q_2 ][(1-\zeta_1)q_2^2+\zeta_2q_1\cdot q_2 ]}{ q_1^2q_2^2\left[\zeta_2q_1^2  +\zeta_1q_2^2 -(\zeta_2q_1-\zeta_1q_2)^2\right]^2}\nonumber\\
 &&+\frac{\Big[(\zeta_1l-q_1)\cdot [(\zeta_1l-q_1)+(\zeta_2q_1-\zeta_1q_2)]\Big]\Big[ (\zeta_2l-q_2)\cdot[(\zeta_2l-q_2)+(\zeta_1q_2-\zeta_2q_1)]\Big]}{\left[(\zeta_1l-q_1)^2\right]\left[(\zeta_2l-q_2)^2\right]\left[\zeta_2(\zeta_1l-q_1)^2  +\zeta_1(\zeta_2l-q_2)^2 -(\zeta_2q_1-\zeta_1q_2)^2\right]^2}\nonumber\\
 &&-\frac{\Big[q_1\cdot[(\zeta_1l-q_1)+(\zeta_2q_1-\zeta_1q_2)] \Big]\Big[ (\zeta_2l-q_2)\cdot[(1-\zeta_1)q_2+\zeta_2q_1]\Big]}{q_1^2\left[(\zeta_2l-q_2)^2\right]\left[\zeta_2q_1^2  +\zeta_1q_2^2 -(\zeta_2q_1-\zeta_1q_2)^2\right]\left[\zeta_2(\zeta_1l-q_1)^2  +\zeta_1(\zeta_2l-q_2)^2 -(\zeta_2q_1-\zeta_1q_2)^2\right]}\nonumber\\
  &&-\frac{\Big[q_2\cdot[(\zeta_2l-q_2)+(\zeta_1q_2-\zeta_2q_1)] \Big]\Big[ (\zeta_1l-q_1)\cdot[(1-\zeta_2)q_1+\zeta_1q_2]\Big]}{q_2^2\left[(\zeta_1l-q_1)^2\right]\left[\zeta_2q_1^2  +\zeta_1q_2^2 -(\zeta_2q_1-\zeta_1q_2)^2\right]\left[\zeta_2(\zeta_1l-q_1)^2  +\zeta_1(\zeta_2l-q_2)^2 -(\zeta_2q_1-\zeta_1q_2)^2\right]}\Bigg\}\nonumber\\
&&+\frac{\Big[q_1\cdot\Big((\zeta_1l-q_1)+(\zeta_2q_1-\zeta_1q_2)\Big)\Big]\Big[q_2\cdot \Big((\zeta_2l-q_2)+(\zeta_1q_2-\zeta_2q_1)\Big)\Big] }{q_1^2q_2^2\Big((\zeta_2l-q_2)+(\zeta_1q_2-\zeta_2q_1)\Big)^2\Big((\zeta_1l-q_1)+(\zeta_2q_1-\zeta_1q_2)\Big)^2}\nonumber\\
&&+\frac{\Big[(\zeta_1l-q_1)\cdot\Big((\zeta_1l-q_1)+(\zeta_2q_1-\zeta_1q_2)\Big)\Big]\Big[(\zeta_2l-q_2)\cdot \Big((\zeta_2l-q_2)+(\zeta_1q_2-\zeta_2q_1)\Big)\Big]}{(\zeta_1l-q_1)^2(\zeta_2l-q_2)^2\Big((\zeta_2l-q_2)+(\zeta_1q_2-\zeta_2q_1)\Big)^2\Big((\zeta_1l-q_1)+(\zeta_2q_1-\zeta_1q_2)\Big)^2}\nonumber\\
&&+\frac{\Big[q_1\cdot\Big((\zeta_1l-q_1)+(\zeta_2q_1-\zeta_1q_2)\Big)\Big]\Big[(\zeta_2l-q_2)\cdot \Big((\zeta_2l-q_2)+(\zeta_1q_2-\zeta_2q_1)\Big)\Big] }{q_1^2(\zeta_2l-q_2)^2\Big((\zeta_2l-q_2)+(\zeta_1q_2-\zeta_2q_1)\Big)^2\Big((\zeta_1l-q_1)+(\zeta_2q_1-\zeta_1q_2)\Big)^2}\nonumber\\
&&+\frac{\Big[(\zeta_1l-q_1)\cdot\Big((\zeta_1l-q_1)+(\zeta_2q_1-\zeta_1q_2)\Big)\Big]\Big[q_2\cdot \Big((\zeta_2l-q_2)+(\zeta_1q_2-\zeta_2q_1)\Big)\Big] }{(\zeta_1l-q_1)^2q_2^2\Big((\zeta_2l-q_2)+(\zeta_1q_2-\zeta_2q_1)\Big)^2\Big((\zeta_1l-q_1)+(\zeta_2q_1-\zeta_1q_2)\Big)^2}\nonumber\\
&&+\Bigg[\frac{\zeta_1}{1-\zeta_1}\Bigg\{-\frac{\Big[q_1\cdot \Big((\zeta_1l-q_1)+(\zeta_2q_1-\zeta_1q_2)\Big)\Big]\Big[q_2\cdot \Big(q_2-(\zeta_1q_2-\zeta_2q_1)\Big)\Big]}{q_1^2q_2^2\left[\zeta_2q_1^2  +\zeta_1q_2^2 -(\zeta_2q_1-\zeta_1q_2)^2\right]\Big((\zeta_1l-q_1)+(\zeta_2q_1-\zeta_1q_2)\Big)^2}\nonumber\\
&&-\frac{\Big[q_1\cdot \Big((\zeta_1l-q_1)+(\zeta_2q_1-\zeta_1q_2)\Big)\Big]\Big[(\zeta_2l-q_2)\cdot \Big(q_2-(\zeta_1q_2-\zeta_2q_1)\Big)\Big]}{q_1^2(\zeta_2l-q_2)^2\left[\zeta_2q_1^2  +\zeta_1q_2^2 -(\zeta_2q_1-\zeta_1q_2)^2\right]\Big((\zeta_1l-q_1)+(\zeta_2q_1-\zeta_1q_2)\Big)^2}\nonumber\\
&&+\frac{\Big[q_2\cdot \Big((\zeta_2l-q_2)+(\zeta_1q_2-\zeta_2q_1)\Big) \Big]\Big[(\zeta_1l-q_1)\cdot \Big((\zeta_1l-q_1)+(\zeta_2q_1-\zeta_1q_2)\Big)\Big]}{q_2^2(\zeta_1l-q_1)^2\Big((\zeta_1l-q_1)+(\zeta_2q_1-\zeta_1q_2)\Big)^2\Big[\zeta_2(\zeta_1l-q_1)^2  +\zeta_1(\zeta_2l-q_2)^2 -(\zeta_2q_1-\zeta_1q_2)^2\Big]}\nonumber\\
&&+\frac{\Big[(\zeta_2l-q_2)\cdot \Big((\zeta_2l-q_2)+(\zeta_1q_2-\zeta_2q_1)\Big) \Big]\Big[(\zeta_1l-q_1)\cdot \Big((\zeta_1l-q_1)+(\zeta_2q_1-\zeta_1q_2)\Big)\Big]}{(\zeta_2l-q_2)^2(\zeta_1l-q_1)^2\Big((\zeta_1l-q_1)+(\zeta_2q_1-\zeta_1q_2)\Big)^2\Big[\zeta_2(\zeta_1l-q_1)^2  +\zeta_1(\zeta_2l-q_2)^2 -(\zeta_2q_1-\zeta_1q_2)^2\Big]}\Bigg\}\nonumber\\
&& +(\zeta_1,q_1\leftrightarrow \zeta_2,q_2)\Bigg].\
 \end{eqnarray}
 \begin{eqnarray}  \label{fn3}
  &&\bar F^2_{12}=-g^4\frac{4\zeta_1\zeta_2}{p^{+2}}\bar\Sigma(12;21)\Bigg[\frac{\zeta_1\zeta_2}{(1-\zeta_1)(1-\zeta_2)}\Bigg\{\frac{\zeta_1\zeta_2\Big[q_1^2q_2^2-(q_1\cdot q_2)^2\Big]}{ q_1^2q_2^2\left[\zeta_2q_1^2  +\zeta_1q_2^2 -(\zeta_2q_1-\zeta_1q_2)^2\right]^2}\nonumber\\
  &&-\frac{\Big[(\zeta_1l-q_1)\times(\zeta_2q_1-\zeta_1q_2)\Big]\Big[ (\zeta_2l-q_2)\times(\zeta_1q_2-\zeta_2q_1)\Big]}{\left[(\zeta_1l-q_1)^2\right]\left[(\zeta_2l-q_2)^2\right]\left[\zeta_2(\zeta_1l-q_1)^2  +\zeta_1(\zeta_2l-q_2)^2 -(\zeta_2q_1-\zeta_1q_2)^2\right]^2}\nonumber\\
 &&+\frac{\Big[q_1\times(\zeta_1l-\zeta_1q_2)\Big]\Big[ (\zeta_2l-q_2)\times[(1-\zeta_1)q_2+\zeta_2q_1]\Big]}{q_1^2\left[(\zeta_2l-q_2)^2\right]\left[\zeta_2q_1^2  +\zeta_1q_2^2 -(\zeta_2q_1-\zeta_1q_2)^2\right]\left[\zeta_2(\zeta_1l-q_1)^2  +\zeta_1(\zeta_2l-q_2)^2 -(\zeta_2q_1-\zeta_1q_2)^2\right]}\nonumber\\
 &&+\frac{\Big[q_2\times(\zeta_2l-\zeta_2q_1) \Big]\Big[ (\zeta_1l-q_1)\times[(1-\zeta_2)q_1+\zeta_1q_2]\Big]}{q_2^2\left[(\zeta_1l-q_1)^2\right]\left[\zeta_2q_1^2  +\zeta_1q_2^2 -(\zeta_2q_1-\zeta_1q_2)^2\right]\left[\zeta_2(\zeta_1l-q_1)^2  +\zeta_1(\zeta_2l-q_2)^2 -(\zeta_2q_1-\zeta_1q_2)^2\right]}\Bigg\}\nonumber\\
  &&-\frac{\Big[q_1\times(\zeta_1l-\zeta_1q_2)\Big]\Big[q_2\times (\zeta_2l-\zeta_2q_1)\Big] }{q_1^2q_2^2\Big((\zeta_2l-q_2)+(\zeta_1q_2-\zeta_2q_1)\Big)^2\Big((\zeta_1l-q_1)+(\zeta_2q_1-\zeta_1q_2)\Big)^2}\nonumber\\
&&-\frac{\Big[(\zeta_1l-q_1)\times(\zeta_2q_1-\zeta_1q_2)\Big]\Big[(\zeta_2l-q_2)\times (\zeta_1q_2-\zeta_2q_1)\Big]}{(\zeta_1l-q_1)^2(\zeta_2l-q_2)^2\Big((\zeta_2l-q_2)+(\zeta_1q_2-\zeta_2q_1)\Big)^2\Big((\zeta_1l-q_1)+(\zeta_2q_1-\zeta_1q_2)\Big)^2}\nonumber\\
&&-\frac{\Big[q_1\times(\zeta_1l-\zeta_1q_2)\Big]\Big[(\zeta_2l-q_2)\times(\zeta_1q_2-\zeta_2q_1)\Big] }{q_1^2(\zeta_2l-q_2)^2\Big((\zeta_2l-q_2)+(\zeta_1q_2-\zeta_2q_1)\Big)^2\Big((\zeta_1l-q_1)+(\zeta_2q_1-\zeta_1q_2)\Big)^2}\nonumber\\
&&-\frac{\Big[(\zeta_1l-q_1)\times(\zeta_2q_1-\zeta_1q_2)\Big]\Big[q_2\times (\zeta_2l-\zeta_2q_1)\Big] }{(\zeta_1l-q_1)^2q_2^2\Big((\zeta_2l-q_2)+(\zeta_1q_2-\zeta_2q_1)\Big)^2\Big((\zeta_1l-q_1)+(\zeta_2q_1-\zeta_1q_2)\Big)^2}\nonumber\\
&&+\Bigg[\frac{\zeta_1}{1-\zeta_1}\Bigg\{\frac{\zeta_2\Big[q_1\times (\zeta_1l-\zeta_1q_2)\Big]\Big[q_2\times q_1\Big]}{q_1^2q_2^2\left[\zeta_2q_1^2  +\zeta_1q_2^2 -(\zeta_2q_1-\zeta_1q_2)^2\right]\Big((\zeta_1l-q_1)+(\zeta_2q_1-\zeta_1q_2)\Big)^2}\nonumber\\
&&+\frac{\Big[q_1\times (\zeta_1l-\zeta_1q_2)\Big]\Big[(\zeta_2l-q_2)\times \Big(q_2-(\zeta_1q_2-\zeta_2q_1)\Big)\Big]}{q_1^2(\zeta_2l-q_2)^2\left[\zeta_2q_1^2  +\zeta_1q_2^2 -(\zeta_2q_1-\zeta_1q_2)^2\right]\Big((\zeta_1l-q_1)+(\zeta_2q_1-\zeta_1q_2)\Big)^2}\nonumber\\
&&-\frac{\Big[q_2\times (\zeta_2l-\zeta_2q_1) \Big]\Big[(\zeta_1l-q_1)\times (\zeta_2q_1-\zeta_1q_2)\Big]}{q_2^2(\zeta_1l-q_1)^2\Big((\zeta_1l-q_1)+(\zeta_2q_1-\zeta_1q_2)\Big)^2\Big[\zeta_2(\zeta_1l-q_1)^2  +\zeta_1(\zeta_2l-q_2)^2 -(\zeta_2q_1-\zeta_1q_2)^2\Big]}\nonumber\\
&&-\frac{\Big[(\zeta_2l-q_2)\times(\zeta_1q_2-\zeta_2q_1) \Big]\Big[(\zeta_1l-q_1)\cdot (\zeta_2q_1-\zeta_1q_2)\Big]}{(\zeta_2l-q_2)^2(\zeta_1l-q_1)^2\Big((\zeta_1l-q_1)+(\zeta_2q_1-\zeta_1q_2)\Big)^2\Big[\zeta_2(\zeta_1l-q_1)^2  +\zeta_1(\zeta_2l-q_2)^2 -(\zeta_2q_1-\zeta_1q_2)^2\Big]}\Bigg\}\nonumber\\
&& +(\zeta_1,q_1\leftrightarrow \zeta_2,q_2)\Bigg].\
 \end{eqnarray}
 Finally we have
 \begin{equation} \label{c-2g-a}
 \frac{d\sigma}{d ^2k_1dk_1^+d^2k_2dk_2^+}=\frac{1}{2} \int_lN(l)\Big[ F_{11}^2+\bar F_{11}^2+F_{12}^2+\bar F_{12}^2+(q_1,\zeta_1\leftrightarrow q_2,\zeta_2)\Big].
 \end{equation}
Note that the partial symmetrization of some terms in Eqs.\,(\ref{fn2},\ref{fn3}) should not be confused with the full symmetrization in the final expersion  \eq{c-2g-a}. Indeed, there is no double counting.

\subsection{The soft (eikonal) limit}
These expressions are fairly complicated, but it is straightforward to take the soft limit where the photons are much less energetic than the quark. This means $\zeta,\zeta_1,\zeta_2\ll 1$.   In this limit we have
\begin{equation}
B(p-q,s,s',q,i)=g\sqrt{\frac{2}{\zeta p^+}}\frac{2q_i}{q^2}
\delta_{ss'},
\end{equation}
\begin{equation}
C(p-q_1-q_2, s,s',q_1,i,q_2,j)=g^2\frac{8}{p^+}\sqrt{\frac{1}{\zeta_1\zeta_2}}\frac{q_{1i}q_{2j}}{q_1^2(q_2^2+q_1^2\frac{\zeta_2}{\zeta_1})}\delta_{ss'}.
\end{equation}
These simple leading soft expressions give vanishing production amplitude. The reason is simply that they do not depend on the transverse quark momentum $p$, and therefore there is a complete cancellation in the amplitude \eq{F}.
Thus we proceed to expand to next order in $\zeta$. We only need to keep the terms which depend on the momentum $p$, as other terms cancel just like with the leading eikonal pieces. Therefore, we obtain at next to leading order, 
\begin{equation}
B(p-q,s,s',q,i)=g\sqrt{\frac{2}{\zeta p^+}}\frac{2}{q^2}\left[q_i-\zeta \left(p_i-\frac{p^2-(p-q)^2}{q^2}q_i\right)\right]\delta_{ss'}=\sqrt{\frac{2}{\zeta p^+}}\frac{2}{q^2}\left[q_i-\zeta \left(p_i-\frac{2p\cdot q}{q^2}q_i\right)\right]\delta_{ss'}.
\end{equation}
\begin{eqnarray}
C(p-q_1-q_2, s,s',q_1,i,q_2,j)&=&g^2\frac{8}{p^+}\sqrt{\frac{1}{\zeta_1\zeta_2}}\frac{1}{q_1^2(q_2^2+q_1^2\frac{\zeta_2}{\zeta_1})} \nonumber\\
&\times&\left[q_{1i}q_{2j}   -\zeta_1\left(p_{i}-\frac{2p\cdot q_1}{q_1^2}q_{1i}\right)q_{2j}-\zeta_2q_{1i}\left(p_{j} -\frac{2p\cdot (q_1+q_2)}{q_2^2+q_1^2\frac{\zeta_2}{\zeta_1}}q_{2j}\right)\right]\delta_{ss'}.\
\end{eqnarray}    
In both of these expressions we have only kept terms proportional to the quark momentum $p$.
To calculate the amplitude $F$ we need
\begin{equation}\left[B(-q_1,s,s",q_1,i)-B(l-q_1,s,s",q_1,i)\right]B(l-q_1-q_2,s",s',q_2,j)
=g^2\frac{8}{p^+}\sqrt{\frac{\zeta_1}{\zeta_2 }}\frac{1}{q_1^2q_2^2}\left(l_i-\frac{2l\cdot q_1}{q_1^2}q_{1i}\right) q_{2j} \delta_{ss'}.
\end{equation}
\begin{eqnarray}
&& C(-q_1-q_2,s,s";q_1,i,q_2,j)-C(l-q_1-q_2,s,s";q_1,i,q_2,j)=\nonumber\\
&&g^2\frac{8}{p^+}\sqrt{\frac{1}{\zeta_1\zeta_2}}\frac{1}{q_1^2(q_2^2+q_1^2\frac{\zeta_2}{\zeta_1})}\left[\zeta_1\left(l_{i}-\frac{2l\cdot q_1}{q_1^2}q_{1i}\right)q_{2j}+\zeta_2q_{1i}\left(l_{j} -\frac{2l\cdot (q_1+q_2)}{q_2^2+q_1^2\frac{\zeta_2}{\zeta_1}}q_{2j}\right)\right]\delta_{ss'}.\
\end{eqnarray}
Thus in the current approximation we obtain, 
\begin{eqnarray}
&&F(0,l,q_1,q_2,ij,s,s')=\nonumber\\
&&g^2\frac{8}{p^+}\sqrt{\frac{1}{\zeta_1\zeta_2}}\frac{1}{q_1^2}\left[\frac{\zeta_1}{q_2^2}\left(l_i-\frac{2l\cdot q_1}{q_1^2}q_{1i}\right) q_{2j} +\frac{\zeta_1}{q_2^2+q_1^2\frac{\zeta_2}{\zeta_1}}\left(l_{i}-\frac{2l\cdot q_1}{q_1^2}q_{1i}\right)q_{2j}+\frac{\zeta_2}{q_2^2+q_1^2\frac{\zeta_2}{\zeta_1}}q_{1i}\left(l_{j} -\frac{2l\cdot (q_1+q_2)}{q_2^2+q_1^2\frac{\zeta_2}{\zeta_1}}q_{2j}\right)\right]\delta_{ss'}.\nonumber\\
\end{eqnarray}
To calculate the cross section we need to symmetrize this expression with respect to the external photon momenta and polarizations.
  The most interesting situation is when the two photons have very different rapidities. If the correlation function exhibits a genuine long range ridge-like structure, like in the di-hadron case,  such a structure should survive in this limit. We therefore take $z_1\ll z_2$.
In this limit we have
\begin{eqnarray}\label{amp1}
&&F(0,l,k_1,k_2,i,j)+F(0,i,k_2,k_1,j,i)=\nonumber\\
&&g^2\frac{8}{p^+}\sqrt{\frac{1}{z_1z_2}}\left[\frac{z_1}{k_1^2k_2^2}\left(l_i-\frac{2l\cdot k_1}{k_1^2}k_{1i}\right) k_{2j} +\frac{z_1}{k_1^4}k_{1i}l_j  +\frac{2z_2}{k_1^2k_2^2}\left(l_j-\frac{2l\cdot k_2}{k_2^2}k_{2j}\right)k_{1i} +\frac{z_1}{k_1^2k_2^2}k_{2j}\left(l_i-\frac{2l\cdot(k_1+k_2)}{k_1^2}k_{1i}\right)\right]\nonumber\\
&&\rightarrow_{z_1\ll z_2}\frac{16}{p^+}\sqrt{\frac{z_2}{z_1}}\frac{1}{k_1^2k_2^2}\left(l_j-\frac{2l\cdot k_2}{k_2^2}k_{2j}\right)k_{1i}.\
\end{eqnarray}
It is now easy to calculate the production cross section by squaring \eq{amp1}, 
\begin{equation}
\frac{d\sigma}{d ^2k_1dk_1^+d^2k_2dk_2^+}=\frac{g^416^2}{p^{+2}}\frac{z_2}{z_1}\frac{1}{k_1^2k_2^4}\int_lN(l) l^2.
\end{equation}
This obviously shows no angular correlations. Thus if the correlations exist, they can only be significant when $\zeta_1$ and $\zeta_2$ are of similar magnitude.

When calculating the cross section in p+p(A) collisions,  the integration over $x_q$ can pick a contribution from $z_2$ which is not very small. We should consider this possibility. Let us therefore relax our approximation and consider $z_1\ll 1$, but $z_2$ not necessarily small.
As a preamble to this we have
\begin{equation}
C(p-q_1-q_2, s,s',q_1,i,q_2,j)_{\zeta_1\ll1}=-g^4\frac{4}{p^+}\sqrt{\frac{\zeta_1}{\zeta_2}}\frac{1}{q_1^2}
\frac{1}{q_1^2(1-\zeta_2)}q_{1i}\big[\zeta_2(p-q_1)_k-q_{2k}\big]\left[\frac{2-\zeta_2}{\zeta_2}\delta_{kj}\delta_{ss'}-i\epsilon_{kj}\sigma^3_{ss'}\right].
\end{equation}
\begin{eqnarray}
C(p-q_1-q_2, s,s',q_1,i,q_2,j)_{\zeta_2\ll 1}&=&g^4\frac{4}{p^+}\sqrt{\frac{\zeta_1}{\zeta_2}}\frac{1}{q_2^2\left[p^2\zeta_1(1-\zeta_1)-(p-q_1)^2\zeta_1-q_1^2(1-\zeta_1)\right]}q_{2j}\big[\zeta_1 p_l-q_{1l}\big] \nonumber\\
&\times& \left[\frac{2-\zeta_1}{\zeta_1}\delta_{li}\delta_{ss'}-i\epsilon_{li}\sigma^3_{ss'}\right]. \
\end{eqnarray}
We now write the symmetrized expression that enters the amplitude
\begin{eqnarray}
&&\Big[C(p-q_1-q_2, s,s',q_1,i,q_2,j)_{q_1=k_1,q_2=k_2,\zeta_1=z_1,\zeta_2=z_2}+C(p-q_1-q_2, s,s',q_1,i,q_2,j)_{q_2=k_1,q_1=k_2;\zeta_2=z_1,\zeta_1=z_2}\Big]_{z_1\ll 1}\nonumber\\
&&=-g^4\frac{4}{p^+}\sqrt{\frac{z_1}{z_2}}\frac{1}{k_1^2}
\frac{1}{k_1^2(1-z_2)}k_{1i}\big[z_2(p-k_1)_k-k_{2k}\big]\left[\frac{2-z_2}{z_2}\delta_{kj}\delta_{ss'}-i\epsilon_{kj}\sigma^3_{ss'}\right]\nonumber\\
&&+\frac{4}{p^+}\sqrt{\frac{z_2}{z_1}}\frac{1}{k_1^2\left[p^2z_2(1-z_2)-(p-k_2)^2z_2-k_2^2(1-z_2)\right]}k_{1i}\big[z_2 p_l-k_{2l}\big]\left[\frac{2-z_2}{z_2}\delta_{lj}\delta_{ss'}-i\epsilon_{lj}\sigma^3_{ss'}\right]\nonumber\\
&&=g^4\frac{4}{p^+}\sqrt{\frac{z_2}{z_1}}\frac{1}{k_1^2\left[p^2z_2(1-z_2)-(p-k_2)^2z_2-k_2^2(1-z_2)\right]}k_{1i}\big[z_2 p_l-k_{2l}\big]\left[\frac{2-z_2}{z_2}\delta_{lj}\delta_{ss'}-i\epsilon_{lj}\sigma^3_{ss'}\right], 
\end{eqnarray}
where the last equality follows since $z_1\ll 1$.

Next consider
\begin{equation}
D(p,l,s,s',q_1,i,q_2,j)\equiv \Big[B(p-q_1,s,s",q_1,i)-B(p+l-q_1,s,s",q_1,i)\Big]B(p+l-q_1-q_2,s",s',q_2,j). 
\end{equation}
We have
\begin{eqnarray}
D(p,l,s,s',q_1,i,q_2,j)_{\zeta_1\ll 1}&=& O(\sqrt \zeta_1)\nonumber\\
D(p,l,s,s',q_1,i,q_2,j)_{\zeta_2\ll 1}&=&g^2\frac{4}{p^+}\sqrt{\frac{\zeta_1}{\zeta_2 }}q_{2j}\left[\frac{2-\zeta_1}{\zeta_1}\delta_{ki}\delta_{ss'}-i\epsilon_{ki}\sigma^3_{ss'}\right]\frac{1}{q_2^2}  \nonumber\\
&\times& \Big[\frac{1}{p^2\zeta_1(1-\zeta_1)-(p-q_1)^2\zeta_1-q_1^2(1-\zeta_1)}(\zeta_1 p_k-q_{1k}) \nonumber\\
&-&\frac{1}{(p+l)^2\zeta_1(1-\zeta_1)-(p+l-q_1)^2\zeta_1-q_1^2(1-\zeta_1)}(\zeta_1 (p_k+l_k)-q_{1k})\Big]. \
\end{eqnarray}
Therefore
\begin{eqnarray}
&&\Big[D(p,l, s,s',q_1,i,q_2,j)_{q_1=k_1,q_2=k_2,\zeta_1=z_1,\zeta_2=z_2}+D(p,l, s,s',q_1,i,q_2,j)_{q_2=k_1,q_1=k_2;\zeta_2=z_1,\zeta_1=z_2}\Big]_{z_1\ll 1}=\nonumber\\
&&g^2\frac{4}{p^+}\sqrt{\frac{z_2}{z_1 }}k_{1i}\left[\frac{2-z_2}{z_2}\delta_{kj}\delta_{ss'}-i\epsilon_{kj}\sigma^3_{ss'}\right]\frac{1}{k_1^2} \Big[
\frac{1}{p^2z_2(1-z_2)-(p-k_2)^2z_2-k_2^2(1-z_2)}(z_2 p_k-k_{2k})\nonumber\\
&-&\frac{1}{(p+l)^2z_2(1-z_2)-(p+l-k_2)^2z_2-k_2^2(1-z_2)}(z_2(p_k+l_k)-k_{2k})\Big].\
\end{eqnarray}
These expressions can be further simplified, since $p=0$ in the collinear factorization approach. This is however not necessary. It is clear from the above expressions that
in this limit the full amplitude $F+F^T$ only depends on the transverse momentum $k_1$ through the factor $k_i/k_1^2$. The cross section therefore again contains no correlations between $k_1$ and $k_2$. This indicates that the correlations may appear only when both $z_1$ and $z_2$ are not too small. In Sec. III, we present our numerical results for the full expression  \eq{c-2g-a} and show that they reproduce special cases considered in this Appendix. 

\subsection{Separating fragmentation and direct parts in a limiting case}
The cross-section of the prompt di-photon can be generally written as a sum of direct and fragmentation contributions.  
Note that for the general case, such an expression  would be very lengthy (see Appendix A), and would not add any new insight, especially since here we stay away from the collinear kinematics. 
However, the general structure of collinear divergence in master equation (\ref{c-2g}), is similar to the simple case of the soft approximation considered in Ref.\,\cite{di-photon}.  The cross-section for production of a quark  with momentum $q$ and two prompt photons with momenta $k_1$ and $k_2$  in the scattering of an on-shell quark with momentum $p$ off a hadronic target, in a kinematic limit that $|k_{1,2}|<|p-q|$ (soft approximation), can be written in the following simple form in terms of fragmentation and direct parts  \cite{di-photon},  
\bea 
&&\frac{d\sigma^{\text{Direct}}}{d^2{\bf b}d^2{\bf k}_{1T} d\eta_{1}d^2{\bf k}_{2T} d\eta_{2} }=\nonumber\\
&&
\frac{2\alpha_{em}^2e^4_q}{(2\pi)^6 } \int_{q_T^2>\mu_F^2}d^2 {\bf q}_T\,  \frac{|{\bf q}_T+{\bf k}_{1T}/\zeta_1|^4}{q_T^2} 
N_F\Big(|{\bf q}_T+{\bf k}_{1T}(1+1/\zeta_1)+{\bf k}_{2T}|, x_ {g}\left(|{\bf q}_T+{\bf k}_{1T}/\zeta_1| \right) \Big) \nonumber\\
&\times&\zeta_2^2\Bigg[\frac{1}{ k_{1T}^2k_{2T}^2\left(q_T^2 \zeta_1^2+\zeta_2^2|{\bf q}_T+{\bf k}_{1T}/\zeta_1-{\bf k}_{2T}/\zeta_2|^2\right)}  
+
\left( \frac{k_1^+k_2^+}{\mathcal{O} k_{2T}^2}+\frac{k_1^{+2}}{\mathcal{O} k_{1T}^2}\right) \frac{1}{k_2^+q_T^2 \zeta_1^2+k_1^+\zeta_2^2|{\bf q}_T+{\bf k}_{1T}/\zeta_1-{\bf k}_{2T}/\zeta_2|^2} \Bigg]\nonumber\\
&+& (k_{1}\leftrightarrow k_{2}, \zeta_{1}\leftrightarrow \zeta_{2}), \label{ff-d0}\
\eea
\bea
&&\frac{d\sigma^{\text{Fragmentation}}}{d^2{\bf b}d^2{\bf k}_{1T} d\eta_{1}d^2{\bf k}_{2T} d\eta_{2} } |_{ \frac{{\bf k}_{1T}}{\zeta_1}\neq \frac{{\bf k}_{2T}}{\zeta_2}} =
 \frac{\alpha_{em}e^2_q}{2(2\pi)^4 } \frac{k_{1T}^2\zeta_2^2}{|{\bf k}_{1T}\zeta_2-{\bf k}_{2T}\zeta_1|^2} \Bigg[ 
\frac{1}{k_{2T}^2}  +\frac{k_{1T}^2k_2^+ }{k_{2T}^2 \mathcal{O}}+\frac{k^+_1}{\mathcal{O}} \Bigg] 
\frac{1}{\zeta_1}D_{\gamma/h}(\zeta_1,\mu_F^2) \nonumber\\
&\times&N_F\Big(|{\bf k}_{1T}(1+1/\zeta_1)+{\bf k}_{2T}|,x_{g}\left(k_{1T}/\zeta_1 \right)\Big)+ (k_{1}\leftrightarrow k_{2}, z_{1}\leftrightarrow \zeta_{2} ), \label{ff-d2}\nonumber\\
&&\frac{d\sigma^{\text{Fragmentation}}}{d^2{\bf b}d^2{\bf k}_{1T} d\eta_{\gamma_1}d^2{\bf k}_{2T} d\eta_{\gamma_2} } |_{ \frac{{\bf k}_{1T}}{\zeta_1} =\frac{{\bf k}_{2T}}{\zeta_2}}=
 \frac{\alpha_{em}e^2_q}{2(2\pi)^4}  \Bigg[ \frac{1}{2} +\left( k_{1}^{+} k_2^+  k_{1T}^2+k_1^{+2} k_{2T}^2\right) \frac{\zeta_2^2}{\mathcal{O}(k^+_2\zeta_1^2+k^+_1\zeta_2^2)}\Bigg] \frac{1}{\zeta_1 \zeta_2} D_{\gamma_1\gamma_2/h}(\zeta_1,\zeta_2, \mu_F^2)
\nonumber\\
&\times & N_F\Big( |{\bf k}_{1T}(1+1/\zeta_1)+{\bf k}_{2T}|,x_{g}\left(k_{1T}/z_{1}\right)\Big)+(k_{1}\leftrightarrow k_{2}, z_{1}\leftrightarrow \zeta_{2}), \label{ff-d3}\
\eea
where $e_q$ is the fractional electric charge of the projectile quark, and the function $\mathcal{O}$ is defined as
\bea 
\mathcal{O}&=&k_{1T}^2k^{+}_2 + k_{2T}^2k^{+}_1,  \label{o-d}\
\eea
with  $k_1^+=\frac{k_{1T}}{\sqrt{2}}e^{\eta_{1}}$ and $k_2^+=\frac{ k_{2T}}{\sqrt{2}}e^{\eta_{2}}$. 
In \eq{ff-d3},  $D_{\gamma/h}$ and $D_{\gamma_1\gamma_2/h}$  are the single photon and double photon fragmentation functions. The fragmentation parameters $\zeta_1$ and $\zeta_2$ are defined in \eq{xqg}.  The fragmentation part was separated from the direct di-photon cross-section  by introducing a hard cutoff, the fragmentation scale $\mu_F$, which defines the collinear singular part that is subsumed in the photon fragmentation contribution.  In \eq{ff-d3}, the first and second term give the single and double photon fragmentation contribution, corresponding to the kinematics where only one photon or both photons is emitted almost collinearly with the outgoing quark.


\end{document}